\documentclass{aa}

\usepackage{graphicx}
\usepackage[usenames,dvipsnames]{xcolor}
\usepackage{natbib,twoopt}
\usepackage[flushleft]{threeparttable}
\usepackage{booktabs}
\usepackage[breaklinks=true,colorlinks=true,linkcolor=NavyBlue!50!Emerald,citecolor=NavyBlue!50!Emerald,filecolor=blue,urlcolor=PineGreen!50!NavyBlue]{hyperref}
\usepackage[T1]{fontenc}
\usepackage{ae,aecompl}
\usepackage{newtxtext}
\usepackage[frenchmath,varg]{newtxmath}
\def\micron{\hbox{$\mu$m}}

\begin{document}

   \title{Measuring Chemical Abundances with Infrared Nebular Lines: \textsc{Hii-Chi-mistry-IR}}
   \titlerunning{Measuring chemical abundances from IR nebular lines}

   \author{J.A. Fern\'andez-Ontiveros \inst{1} \and E. P\'erez-Montero\inst{2} \and J.M. V\'ilchez\inst{2} \and R. Amor\'in\inst{3,4} \and L. Spinoglio\inst{1}
          }
   \authorrunning{Fern\'andez-Ontiveros et al.}

   \institute{Istituto di Astrofisica e Planetologia Spaziali (INAF--IAPS), Via Fosso del Cavaliere 100, I--00133 Roma, Italy\\
   \email{\sf \href{mailto:j.a.fernandez.ontiveros@gmail.com}{j.a.fernandez.ontiveros@gmail.com}}
   \and
   Instituto de Astrof\'isica de Andaluc\'ia (IAA-CSIC), Glorieta de la Astronom\'ia s/n, E--18008 Granada, Spain\\
   \email{\sf \href{mailto:epm@iaa.es}{epm@iaa.es}}
   \and
   Instituto de Investigaci\'on Multidisciplinar en Ciencia y Tecnolog\'ia, Universidad de La Serena, Ra\'ul Bitr\'an 1305, La Serena, Chile
   \and
   Departamento de Astronom\'ia, Universidad de La Serena, Av. Juan Cisternas 1200 Norte, La Serena, Chile
             }

   \date{\today}

   \abstract{In this work we provide a new method to derive heavy element abundances taking advantage of the unique suite of fine-structure lines in the mid- to far-infrared (IR) range. Using grids of photo-ionisation models that cover a wide range in O/H and N/O abundances, and ionisation parameter, our code \textsc{Hii-Chi-mistry-IR (\textsc{HCm-IR})} provides model-based abundances based on extinction free and temperature insensitive tracers, two significant advantages over optical diagnostics when these are applied to, e.g. dust obscured regions or unresolved (stratified) nebulae, typical conditions found in high-$z$ galaxies. The performance of the code is probed using three different samples of galaxies extended over a wide range in metallicity, $7.2 \lesssim 12 + \log(O/H) \lesssim 8.9$, with available mid- to far-IR spectroscopic observations from \textit{Spitzer} and \textit{Herschel}, respectively. These correspond to 28 low-metallicity dwarf galaxies, 19 nearby starbursts, and 9 luminous IR galaxies. The IR model-based metallicities obtained are robust within a scatter of $0.03\, \rm{dex}$ when the hydrogen recombination lines, which are typically faint transitions in the IR range, are not available. When compared to the optical abundances obtained with the direct method, model-based methods, and strong-line calibrations, \textsc{HCm-IR} estimates show a typical dispersion of $\sim 0.2\, \rm{dex}$, in line with previous studies comparing IR and optical abundances, a do not introduce a noticeable systematic above $12 + \log(O/H) \gtrsim 7.6$. This accuracy can be achieved by measuring the sulphur ([\ion{S}{iv}]$_{\rm 10.5 \mu m}$ and [\ion{S}{iii}]$_{\rm 18.7,33.5 \mu m}$) and the neon ([\ion{Ne}{iii}]$_{\rm 15.6 \mu m}$ and [\ion{Ne}{ii}]$_{\rm 12.8 \mu m}$) lines. Additionally, \textsc{HCm-IR} provides an independent N/O measurement when the oxygen ([\ion{O}{iii}]$_{\rm 52,88 \mu m}$) and nitrogen ([\ion{N}{iii}]$_{\rm 57 \mu m}$) transitions are measured, and therefore the derived abundances in this case do not rely on particular assumptions in the N/O ratio. Large uncertainties ($\sim 0.4\, \rm{dex}$) may affect the abundance determinations of galaxies at sub- or over-solar metallicities when a solar-like N/O ratio is adopted. Finally, the code has been applied to 8 galaxies located at $1.8 < z < 7.5$ with ground-based detections of far-IR lines redshifted in the submillimetre range, revealing solar-like N/O and O/H abundances in agreement with recent studies. A script to derive chemical abundances with \textsc{HCm-IR} has been made publicly available online.}


   \maketitle
%

\section{Introduction}\label{intro}

Elements heavier than hydrogen and helium are known as metals and represent a very small fraction of the total baryon mass in the present Universe ($\sim 0.02\%$; \citealt{madau2014,maiolino2019}), however they play a fundamental role in essentially every astrophysical environment from the evolution of galaxies till the formation of stellar systems and planets, and the emergence of life. Understanding these processes requires reliable diagnostics and robust tracers to measure the abundances of the different chemical elements.

The first witness of the chemical enrichment is the interstellar medium (ISM), where heavy elements are returned at the end of the stellar evolution. Gas-phase metallicities in star-forming regions, planetary nebulae, and the ISM of galaxies are typically measured using nebular lines in their optical spectra (see \citealt{peimbert2017,maiolino2019} for reviews). The direct method provides accurate metallicities based on collisionally excited lines of heavy elements, whose emissivities are known once the density and temperature of the gas are determined \citep[e.g.][]{izotov1994,izotov1997,izotov1999,kennicutt2003,haegele2008}. The latter is a critical step since it depends on the measurement of auroral lines (e.g. [\ion{O}{iii}] $\lambda 4363$), associated with high energy levels, which are typically faint at solar-like metallicities \citep{bresolin2005,moustakas2010}. When these are not available the estimates rely on indirect methods \citep{fernandezmartin2017}, based on ratios of strong nebular lines that are calibrated against determinations based on the direct method and/or photo-ionisation models \citep[e.g.][]{pagel1979,mcgaugh1991,pilyugin2001,pilyugin2005,perezmontero2005}. However, the systematic discrepancies among the different strong-line calibrations can be quite large ($\sim 0.7\, \rm{dex}$; \citealt{kewley2008}), smearing out the results obtained when different line tracers are combined.

\begin{table*}
    \footnotesize
    \setlength{\tabcolsep}{2.pt}
	\centering
	\caption{IR line fluxes for the sample of star forming galaxies used in this work. For each galaxy we provide the name, coordinates, redshift, spectral type (dwarf galaxy, starburst, or ULIRG), the mid-IR line fluxes in units of $10^{-17}\, \rm{W\,m^{-2}}$, and the references in the literature where these measurements were compiled. The complete version of the table is published in the online version of this paper.}
	\label{tab_sample}
	\resizebox{\textwidth}{!}{
	\begin{threeparttable}[b]
	\begin{tabular}{lccccccccccccc} 
		\bf Name & \bf R.A. (J2000) & \bf Dec. (J2000) & \bf z & \bf Type & {\bf [\ion{S}{iv}]}$10.5\, \rm{\micron}$ & \bf Hu-$\alpha$ & {\bf [\ion{Ne}{ii}]}$12.8\, \rm{\micron}$ & {\bf [\ion{Ne}{iii}]}$15.6\, \rm{\micron}$ & {\bf [\ion{O}{iii}]}$52\, \rm{\micron}$ & {\bf [\ion{N}{iii}]}$57\, \rm{\micron}$ & {\bf [\ion{O}{iii}]}$88\, \rm{\micron}$ & {\bf [\ion{N}{ii}]}$122\, \rm{\micron}$ & \bf Refs.\\
		& (hh:mm:ss) & (dd:mm:ss) & & & & & & & & & & & \\
		\hline\\[-0.2cm]
Haro\,11 & 00:36:52.45 & -33:33:16.77 & 0.020598 & Dwarf & $49.4 \pm 1.1$ & $1.6 \pm 0.5$ & $32.7 \pm 0.9$ & $112 \pm 5$ & -- & $28.3 \pm 0.8$ & $172 \pm 3$ & $3.5 \pm 0.3$ & 1 \\
IRAS00397-1312 & 00:42:15.50 & -12:56:03.0 & 0.261717 & ULIRG & $0.27 \pm 0.07$ & -- & $3.78 \pm 0.04$ & $2.1 \pm 0.2$ & $3.0 \pm 0.4$ & -- & -- & -- & 2, 3 \\
NGC\,253 & 00:47:33.07 & -25:17:19.0 & 0.000811 & SB & -- & -- & $2830 \pm 60$ & $200 \pm 10$ & -- & $400 \pm 100$ & $330 \pm 50$ & $410 \pm 40$ & 4 \\
HS0052+2536 & 00:54:56.36 & +25:53:08.0 & 0.045385 & Dwarf & $1.4 \pm 0.3$ & -- & $0.9 \pm 0.1$ & $1.5 \pm 0.5$ & -- & -- & $7.1 \pm 0.4$ & -- & 1 \\
UM311 & 01:15:34.40 & -00:51:46.1 & 0.005586 & Dwarf & $3.7 \pm 0.3$ & -- & $4 \pm 1$ & $10.3 \pm 0.5$ & -- & -- & $51 \pm 3$ & -- & 1 \\
NGC\,625 & 01:35:05.16 & -41:26:08.8 & 0.001321 & Dwarf & $57.2 \pm 0.7$ & $2.4 \pm 0.2$ & $16.6 \pm 0.5$ & $75 \pm 3$ & -- & -- & $245 \pm 3$ & -- & 1 \\
NGC\,891 & 02:22:32.85 & +42:20:52.7 & 0.001761 & SB & -- & $0.7 \pm 0.3$ & $39.4 \pm 0.4$ & $4.5 \pm 0.2$ & -- & -- & -- & $19.4 \pm 0.8$ & 5, 6 \\
NGC\,1140 & 02:54:33.53 & -10:01:42.1 & 0.005007 & Dwarf & $18 \pm 1$ & $0.5 \pm 0.1$ & $18.3 \pm 0.9$ & $63 \pm 4$ & -- & -- & $108 \pm 3$ & $2.1 \pm 0.3$ & 1 \\
NGC\,1222 & 03:08:56.74 & -02:57:18.6 & 0.008079 & SB & $22.2 \pm 0.5$ & $2.1 \pm 0.2$ & $81 \pm 1$ & $89 \pm 2$ & -- & -- & -- & $5.8 \pm 0.7$ & 4 \\
SBS\,0335-052 & 03:37:44.06 & -05:02:40.19 & 0.013519 & Dwarf & $1.48 \pm 0.04$ & -- & $0.07 \pm 0.01$ & $1.24 \pm 0.07$ & -- & -- & $4.4 \pm 0.3$ & -- & 1 \\[0.1cm]
		\hline
	\end{tabular}
	\begin{tablenotes}
	\item (1) \citealt{cormier2015}; (2) \citealt{veilleux2009}; (3) \citealt{pereira-santaella2017}; (4) \citealt{bernard-salas2009}; (5) \citealt{goulding2009}; (6) \citealt{jafo2016}.
    \end{tablenotes}
\end{threeparttable}
}
\end{table*}

\begin{figure}
  \centering
  \includegraphics[width=\columnwidth]{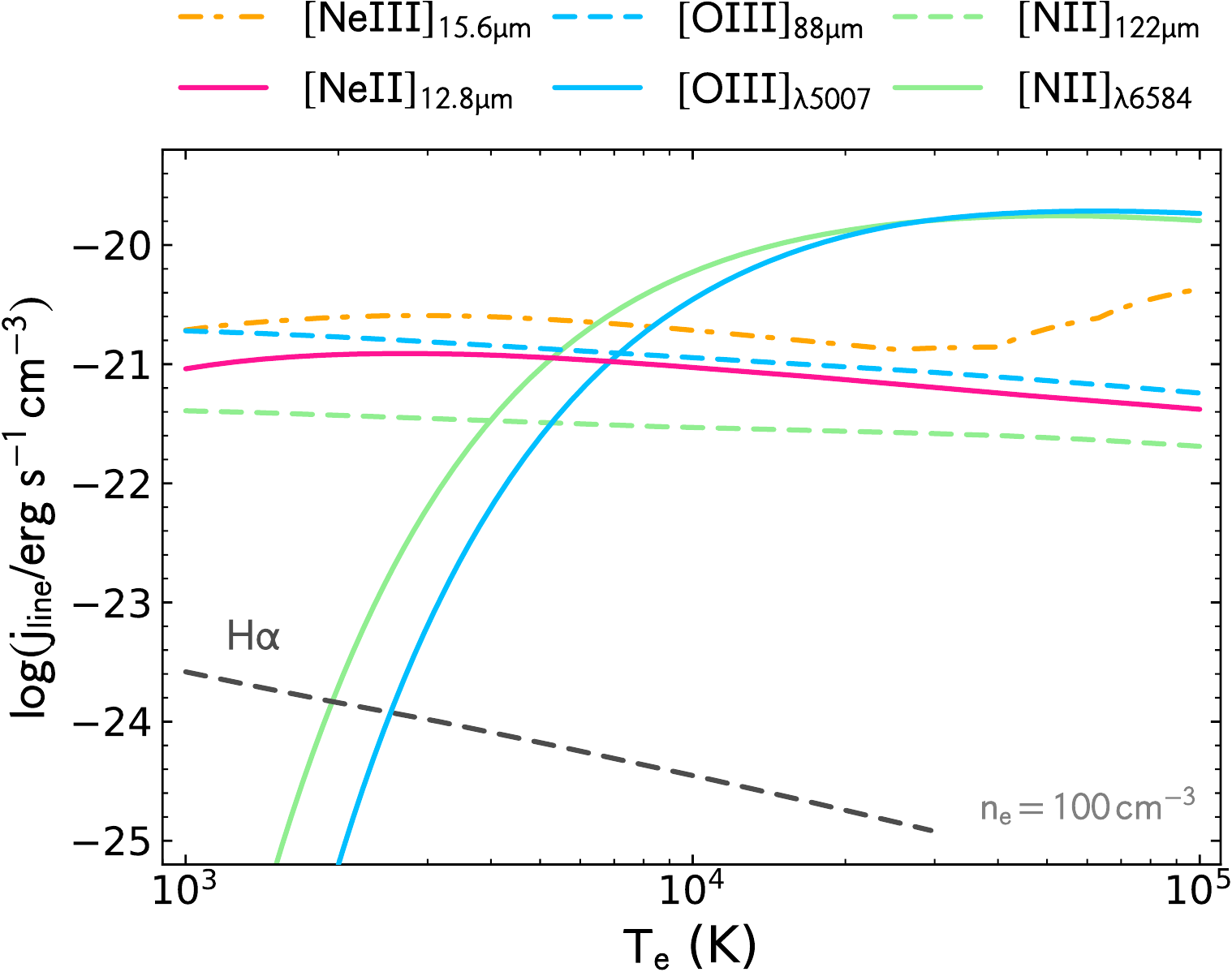}~
  \caption{The emissivity ($j_{\rm line}$) of the fine-structure lines in the optical range (blue solid line: [\ion{O}{iii}] $\lambda 5007$; green solid: [\ion{N}{ii}] $\lambda 6584$) is a strong function of the electron temperature $T_{\rm e}$ in the $2\,000$--$30\,000\, \rm{K}$ range, in contrast with the IR lines (pink solid: [\ion{Ne}{ii}]$12.8\, \rm{\micron}$; orange dot-dashed: [\ion{Ne}{iii}]$15.6\, \rm{\micron}$; blue dashed: [\ion{O}{iii}]$88\, \rm{\micron}$; green dashed: [\ion{N}{ii}]$122\, \rm{\micron}$). The emissivities are computed using \textsc{PyNeb} \citep{luridiana2015} for a fixed $n_{\rm e} = 100\, \rm{cm^{-3}}$. For comparison, the emissivity of the H$\alpha$ recombination line is also shown (black dashed line).}\label{fig_emiss}%
\end{figure}

In the last 30 years the infrared (IR) range has been unfolded by various generations of infrared (IR) spectroscopic observatories, from the \textit{Infrared Space Observatory} (ISO, $2.4$--$197\, \rm{\micron}$; \citealt{kessler1996}) to the \textit{Spitzer Space Telescope} ($5$--$39\, \rm{\micron}$; \citealt{werner2004}) and the \textit{Herschel Space Observatory} ($51$--$671\, \rm{\micron}$; \citealt{pilbratt2010}), providing access to a unique suite of fine-structure lines that serve as diagnostics for a very wide range of physical conditions \citep{spinoglio1992}. In particular, metallicity determinations based on IR lines provide major advantages with respect to the optical and UV estimates \citep[e.g.][]{pottasch1999,verma2003,bernard-salas2008,croxall2013}. First, the IR range is insensitive to the interstellar reddening, thus only IR tracers can probe the gas-phase elemental abundances in the ISM of dusty galaxies \citep{nagao2011,pereira-santaella2017,herrera-camus2018}. This is particularly critical for studies at high-$z$, where $\sim 90\%$ of the light emitted by star forming galaxies is absorbed and reprocessed in the IR \citep{madau2014,santini2010,rowlands2014}, but also in the dusty regions of nearby galaxies where new stars are being formed and thus provide the most recent estimate for the chemical age in these galaxies \citep{verma2003,wu2008}, including our own Galaxy \citep{carr2000,inno2019}. Generally, the nebular emission in galaxies tend to suffer from significant extinction above star formation rates of $\gtrsim 20\, \rm{M_\odot\, yr^{-1}}$ and the gas tend to be selectively affected over the continuum for sub-solar metallicities ($12 + \log(O/H) \lesssim 8.5$; \citealt{reddy2015,shivaei2020}). Second, the emissivity of IR lines shows a feeble dependency with the gas temperature, since the atomic levels involved in the transitions are much closer to the ground state when compared to optical and UV lines \citep[e.g.][]{bernard-salas2001}. In this regard, Fig.\,\ref{fig_emiss} shows that the emissivity of the [\ion{Ne}{ii}]$_{\rm 12.8 \mu m}$ and [\ion{Ne}{iii}]$_{\rm 15.6 \mu m}$ lines change by a factor $\lesssim 4$ in the $1\,000 < T_{\rm e} < 100\,000\, \rm{K}$ range (at $n_{\rm e} = 100\, \rm{cm^{-3}}$; \citealt{luridiana2015}). In contrast, the emissivity of the optical nebular lines varies by five orders of magnitude in the $2\,000 < T_{\rm e} < 30\,000\, \rm{K}$ range. This implies that uncertainties in the temperature determination have a minor effect in the derived IR abundances, including temperature fluctuations in the nebula due to inhomogeneities in the density structure and/or the radiation field that are not spatially resolved in the observed spectra. This is known as the $t^2$ or temperature problem, which may cause a systematic underestimation of the chemical abundances derived from optical lines using the direct method \citep{peimbert1967}, leading to values a factor of $\sim 4$ ($0.6\, \rm{dex}$) lower when compared to IR abundance determinations \citep{vermeij2002,dors2013}. Furthermore, the electron temperature might vary among the various elements and ionic species \citep{garnett1992}, adding a differential effect on the emissivity of the nebular lines that will be stronger in the optical transitions when compared to those in the IR. Additionally, the optical collisionally excited lines cannot detect the contribution from heavy elements in cold plasmas ($\sim 1\,000\, \rm{K}$; see Fig.\,\ref{fig_emiss}), which is one of the possible causes of the abundance discrepancy factors \citep[e.g.][]{liu2006,tsamis2008}.

In this work we extend the methodology developed in the \textsc{Hii-Chi-mistry} code for the optical \citet{perezmontero2014} and the UV lines \citep{perezmontero2017} to determine chemical abundances using the nebular lines in the mid- and far-IR spectral range. The IR line fluxes are compared with grids of photo-ionisation models covering a wide range in N/O and O/H abundances, and ionisation parameter ($\log U$). When applied to optical lines, \textsc{Hii-Chi-mistry} avoids the offsets and systematics that typically affect the strong-line calibrations \citep{perezmontero2014}, providing abundance determinations that are consistent with the direct method with an average offset below $0.1\, \rm{dex}$, that is when the code estimates are compared to those based on a previous determination of $T_{\rm e}$. In Section\,\ref{data} we present a spectral database of galaxies that are used as a case study to test the validity of the new IR-based method. These data include star forming and luminous IR galaxies in the local Universe, and a sample of galaxies at high redshift where far-IR lines have been used to measure the chemical abundances. The IR method is described in detail in Section\,\ref{model}, including the main line ratios used to constrain the N/O and O/H abundances and $\log U$. In Section\,\ref{results} we present the main results obtained with the IR method, which are then discussed in Section\,\ref{discuss}. Finally, the main conclusions are summarised in Section\,\ref{sum}.


\section{Data}\label{data}

As a case study to test the validity of the abundances derived from IR nebular lines, we applied our new method to a sample of star forming galaxies with spectroscopic observations in the mid- to far-IR range. The latter were acquired with the InfraRed Spectrograph (IRS; \citealt{houck2004}) onboard \textit{Spitzer} and the Photodetector Array Camera and Spectrometer (PACS; \citealt{poglitsch2010}) onboard \textit{Herschel}, respectively. A few detections of the [\ion{O}{iii}]$_{\rm 52 \mu m}$ and [\ion{N}{iii}]$_{\rm 57 \mu m}$ far-IR lines obtained with FIFI-LS
\citep{fischer2018} onboard the \textit{SOFIA airborne observatory} \citep{temi2018} were also taken from \citet{peng2021}. The final sample includes 28 dwarf galaxies with sub-solar metallicities \citep{madden2013,cormier2015}, 19 active star forming galaxies with solar-like metallicities \citep{jafo2016}, and 9 low-redshift luminous infrared galaxies (LIRGs and ULIRGs) dominated by the star formation component \citep{pereira-santaella2017,jafo2016}. The AGN contribution in these (U)LIRGs is lower than $10\%$, according to their [\ion{Ne}{v}]$_{\rm 14.3 \mu m}$ / [\ion{Ne}{ii}]$_{\rm 12.8 \mu m}$ measured ratios \citep{pereira-santaella2017}. In all cases we favoured the \textit{Spitzer}/IRS high-spectral resolution mode ($R \sim 600$) when available. Since hydrogen recombination line fluxes are not provided in these works, we completed our spectroscopic sample by measuring the \ion{H}{i} (7-6) Humphreys-$\alpha$ line at $12.37\, \rm{\mu m}$ directly from the \textit{Spitzer} calibrated and extracted spectra available in the CASSIS\footnote{Combined Atlas of Sources with Spitzer IRS Spectra:\\ \url{https://cassis.sirtf.com}} database \citep{lebouteiller2015}. The line flux was obtained by integrating the continuum-subtracted spectrum, and then normalised to the [\ion{Ne}{ii}] $12.81\, \rm{\mu m}$ line flux reported in the literature. Additionally we include flux measurements of the Brackett-$\alpha$ $4.05\, \rm{\mu m}$ line for three of the (U)LIRG galaxies provided by \citet{imanishi2010} with \textit{AKARI}/IRC. The compilation of the IR line fluxes for the sample of galaxies used in this work is provided in Table\,\ref{tab_sample}.

To compare the IR O/H derived abundances with the estimates based on the optical lines we performed a comprehensive search in the literature to compile line fluxes in this range and/or abundance measurements (see Table\,\ref{tab_sample}). For those cases with available line fluxes, the optical metallicities were obtained using the {\sc Hii-Chi-mistry} v4.1 code \citep{perezmontero2014,perezmontero2019}, which relies on the same photo-ionisation models used in this work to derive the IR abundance estimates, described in Section\,\ref{model}.

Finally, we compiled recent measurements for galaxies at high-$z$ ($z \gtrsim 4$). These are sub-millimetre (sub-mm) and Ly$\alpha$ systems where strongly redshifted emission lines of [\ion{O}{iii}]$_{\rm 52,88 \mu m}$ and [\ion{N}{ii}]$_{\rm 122,205 \mu m}$ or [\ion{N}{iii}]$_{\rm 57 \mu m}$ have been detected using observations in the sub-mm range (Table\,\ref{tab_sample}). These results are relevant for chemical evolution studies, and therefore IR based abundances for O/H or N/O in these galaxies are provided in this work.

\section{Model-based abundances}\label{model}

Chemical abundances from IR lines were calculated using an adapted version of the code {\sc Hii-Chi-mistry}, that we denote here as {\sc Hii-Chi-mistry-IR}\footnote{All versions of the {\sc Hii-Chi-mistry} code are publicly available at: \url{http://www.iaa.csic.es/~epm/HII-CHI-mistry.html}.} (hereinafter \textsc{HCm} and \textsc{HCm-IR}, respectively). Both \textsc{HCm} and \textsc{HCm-IR} are based on the same grid of photo-ionisation models. A detailed analysis of the model-based metalicities derived from the optical lines for 550 star-forming emission-line objects in the Local Universe, compared to the measurements obtained with the direct method, show that \textsc{HCm} has a precision better than $0.1\, \rm{dex}$ and do not introduce offsets or systematics with respect to the direct method determinations \citep{perezmontero2014}. Along this Section we describe the specific procedure to determine abundances using {\sc HCm-IR}, however we refer to \citet{perezmontero2014} for a comprehensive description of the models and the general methodology.

This code performs a bayesian-like calculation of the total oxygen abundance, the nitrogen-to-oxygen abundance ratio and the ionisation parameter by comparing a set of reddening-corrected emission-line fluxes with the predictions from a large grid of photo-ionisation models. For this version of the code we used the grid of models computed with {\sc Cloudy} v17.01 {\citep{ferland2017}, adopting the spectral energy distribution (SED) of simple stellar population models from {\sc Popstar} \citep{popstar}, also described in \cite{perezmontero2014}. The grid covers a wide range in oxygen abundance, from $12 + \log(O/H) = 6.9$ to $9.1$ in bins of $0.1\, \rm{dex}$, in nitrogen abundance from $\log(N/O) = -2.0$ to $0.0$ in bins of $0.125\, \rm{dex}$, and in ionisation parameter from $\log U = -4.0$ to $-1.5$ in bins of $0.25\, \rm{dex}$. This yields a total of 4\,301 models, although the code also allows to interpolate between the computed values to multiply the resolution of the grid by a factor 10 in each of the three explored dimensions. All models were calculated assuming a constant electron density of 100 particles per cm$^{-3}$, a filling factor of 0.1, a standard dust-to-gas mass ratio and a plane-parallel geometry. The calculations were stopped in each model when the density of free electrons in relation to hydrogen atoms was lower than 98\%. All species, with the exception of N, were scaled to the solar proportions given by \cite{asplund2009}, and the metallicity of the stars were assumed to match that of the gas in each model.

As input, the \textsc{HCm-IR} code admits reddening-corrected fluxes in arbitrary units --\,with their corresponding errors\,-- of the following emission lines: \ion{H}{i} $4.05\, \rm{\micron}$ (Brackett-$\alpha$ or Br$\alpha$), \ion{H}{i} $7.46\, \rm{\micron}$ (Pfund-$\alpha$ or Pf$\alpha$), [\ion{S}{iv}] $10.5\, \rm{\micron}$, \ion{H}{i} $12.37\, \rm{\micron}$ (Humphreys-$\alpha$ or Hu$\alpha$), [\ion{Ne}{ii}] $12.8\, \rm{\micron}$, [\ion{Ne}{iii}] $15.6\, \rm{\micron}$, [\ion{S}{iii}] $18.7\, \rm{\micron}$, [\ion{S}{iii}] $33.5\, \rm{\micron}$, [\ion{O}{iii}] $52\, \rm{\micron}$, [\ion{N}{iii}] $57\, \rm{\micron}$, [\ion{O}{iii}] $88\, \rm{\micron}$, and [\ion{N}{ii}] $122\, \rm{\micron}$. We did not considered other fine-structure IR lines, such as [\ion{C}{ii}] $157\, \rm{\micron}$ or [\ion{O}{i}] $63\, \rm{\micron}$, which can be partially emitted by the photo-dissociation region (PDR) and are not well reproduced by our models. 
If the errors are also introduced the code can iterate by randomly perturbing the nominal value around the dispersion in a Monte-Carlo simulation to produce realistic uncertainties for the derived chemical abundances and $U$. A significant advantage of this method is that the abundances can be derived even if one or several of the input lines are missing, based on a limited set of input emission-lines. This is especially convenient to compare results for objects observed at different spectral ranges or at different redshifts or for lines whose measurement is not reliable.

The code calculates N/O, O/H, and $\log U$ as the weighted-means of all the input values in all the models of the grid where the weights ($1 / \chi^2$) are obtained as the quadratic sum of the difference between certain predicted and measured emission-line ratios that correlate with one or more of the searched properties.

Additionally, metallicities based on the intensities of optical emission lines were compiled from the literature when available, or derived otherwise from published optical line fluxes (Table\,\ref{tab_abund}) using \textsc{HCm} v4.1, which is based on the same photo-ionisation models as \textsc{HCm-IR}. The lines typically used for the optical abundance determinations are [\ion{O}{ii}] $\lambda 3727$, [\ion{Ne}{iii}] $\lambda 3869$, [\ion{O}{iii}] $\lambda 4363$, [\ion{O}{iii}] $\lambda 5007$, [\ion{N}{ii}] $\lambda 6584$, [\ion{S}{ii}] $\lambda \lambda 6716, 6731$. Figs.\,\ref{n3o3} to \ref{O3N2} show the comparison of the O/H and $\log U$ values derived from the optical lines against the different IR parameters for our sample of local star-forming galaxies (Table\,\ref{tab_sample}).

\begin{figure}
  \centering
  \includegraphics[width=\columnwidth]{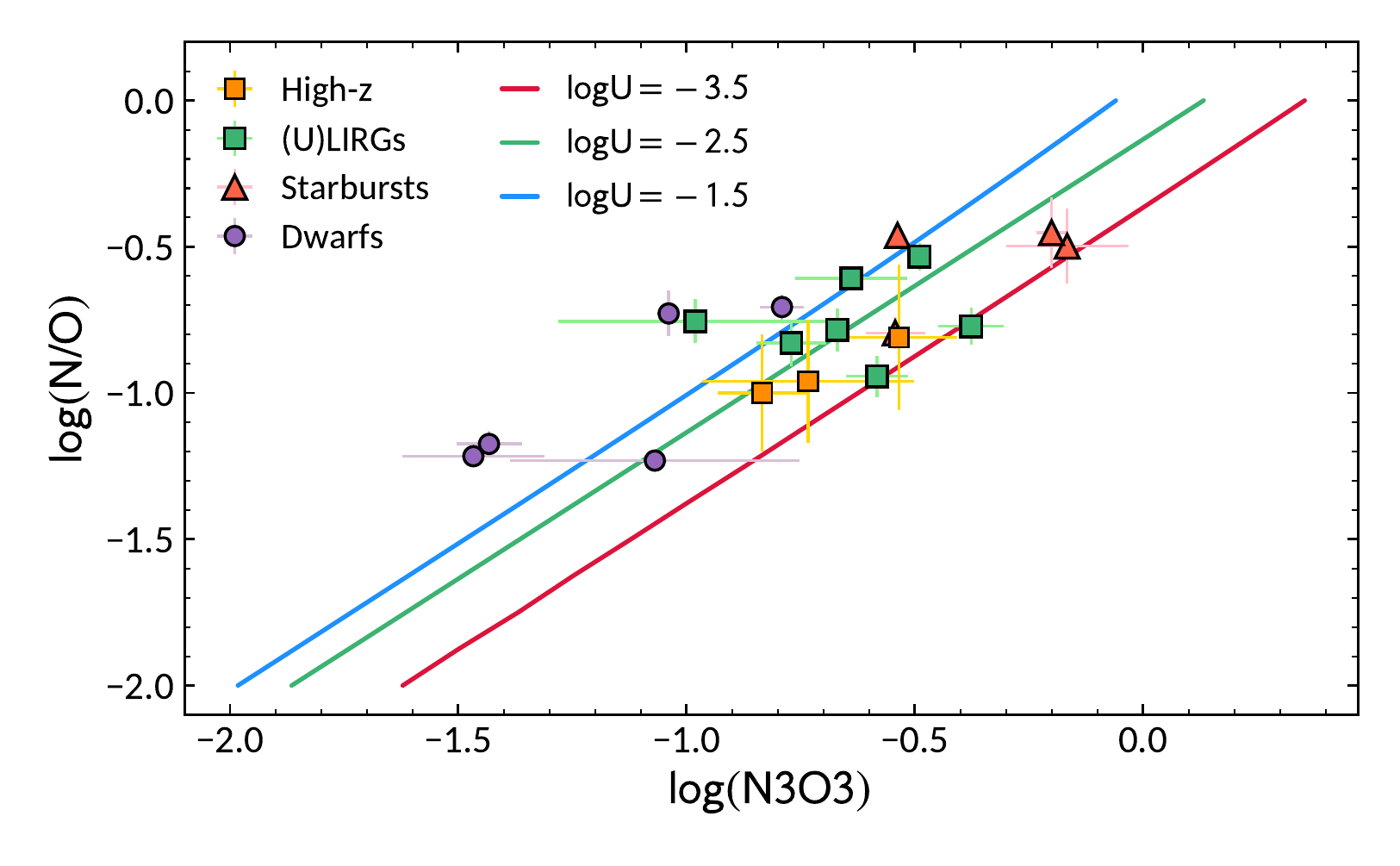}
  \caption{Relation between the N3O3 parameter and the abundance ratio N/O for different models at a fixed $12 + \log(O/H) = 8.9$. The models are compared with the N/O abundances derived from the optical lines for a sample of local star-forming galaxies with IR spectroscopic observations, including dwarf galaxies (purple circles), nearby starbursts (red triangles) and local (U)LIRGs (green squares). Three galaxies at high-$z$ with detected [\ion{N}{iii}] $57\, \rm{\micron}$ and [\ion{O}{iii}] $52, 88\, \rm{\micron}$ are also included (orange squares).}\label{n3o3}%
\end{figure}

\subsection{N/O derivation}\label{NO-derivation}
In a first iteration, the code calculates N/O taking advantage that its derivation is independent of the excitation of the gas. This is equivalent to the case of the optical emission-lines, where certain emission-line ratios, such as N2O2 or N2S2 do not show a dependence on $U$ \citep{pmc09}. In the case of the IR spectral range, a very similar situation is presented with [N{\sc iii}] and [O{\sc iii}] lines, defining the N3O3 parameter:
\begin{equation}\label{eq_n3o3}
{\rm N3O3} =  \log \left( \frac{\rm I([\ion{N}{iii}]_{57 \mu{\rm m}})}{\rm I([\ion{O}{iii}]_{52 \mu{\rm m}}) + I ([\ion{O}{iii}]_{88 \mu{\rm m}})} \right)
\end{equation} 

This parameter has been also proposed by \cite{nagao2011} and \cite{pereira-santaella2017} to provide a direct estimation of the total metallicity. As in the case of N2O2 in the optical \citep{kd02}, this is based on the assumption that for oxygen abundances $12 + \log(O/H) > 8.0$ most of N has a secondary origin and thus N/O increases for increasing O/H. However, this assumption does not hold when hydro-dynamical processes (e.g. inflows/outflows; \citealt{edmunds90}) or variations in the star formation efficiency \citep{molla06} take place, so the relative ratio of a secondary species to a primary one can undergo deviations from the case of a closed-box, as in the case of the so-called \textit{green-pea} galaxies \citep{amorin10}.

As also suggested by \cite{maiolino2019}, N3O3 presents a tight correlation with N/O, with very little dependence on $U$. This linear behaviour is shown in Fig.\,\ref{n3o3} for the model subgrid with a fixed $12 + \log(O/H) = 8.9$, and it is consistent with the observed trend in galaxies. We note that the N/O values shown in Fig.\,\ref{n3o3} for local star-forming galaxies (purple circles: dwarf galaxies; red triangles: nearby starbursts; green squares: local LIRGs) are estimated using \textsc{HCm} from the optical line fluxes of these galaxies, and therefore are derived independently of the measured IR line ratio represented in the horizontal axis. Three galaxies at high-$z$ with IR measurements are also included (orange squares), although the N/O values in these cases are derived with \textsc{HCm-IR} from the IR lines. We show in Fig.\,\ref{n3o3} some of the model-sequences described above (solid lines and coloured circles) to illustrate the clear correlation between N3O3 and N/O and the feeble dependence of this parameter with $U$.
Overall, the N/O values derived from the optical lines for local star-forming galaxies are in agreement with the values of N3O3 measured from the IR lines for the expected range in $U$. An additional contribution to the dispersion not shown in Fig.\,\ref{n3o3} is the spread in metallicity, since all the models shown correspond to $12 + \log(O/H) = 8.9$. The comparison between optical and IR estimates for the different diagnostics shown along this section will be further discussed in Section\,\ref{IR_opt}.

\subsection{O/H and $U$ derivation}\label{OH-derivation}
Once N/O is derived, the code makes a second iteration --\,within the uncertainties\,-- through the subgrid of models compatible with the adopted N/O solution. This ensures that nitrogen lines can be used to derive O/H without any specific assumption on the \mbox{O/H--N/O} relation. However, if N/O cannot be derived within the first iteration, the code assumes a certain relation between O/H and N/O. At low metalicities, i.e. $12 + \log(O/H) < 8.0$, a constant N/O value is adopted since only the primary production of N is considered in star forming galaxies. For higher metallicities, a linear relation between O/H and N/O is used, excluding models that do not follow this trend in the calculation.

\begin{figure*}
  \centering
  \includegraphics[width=0.5\textwidth]{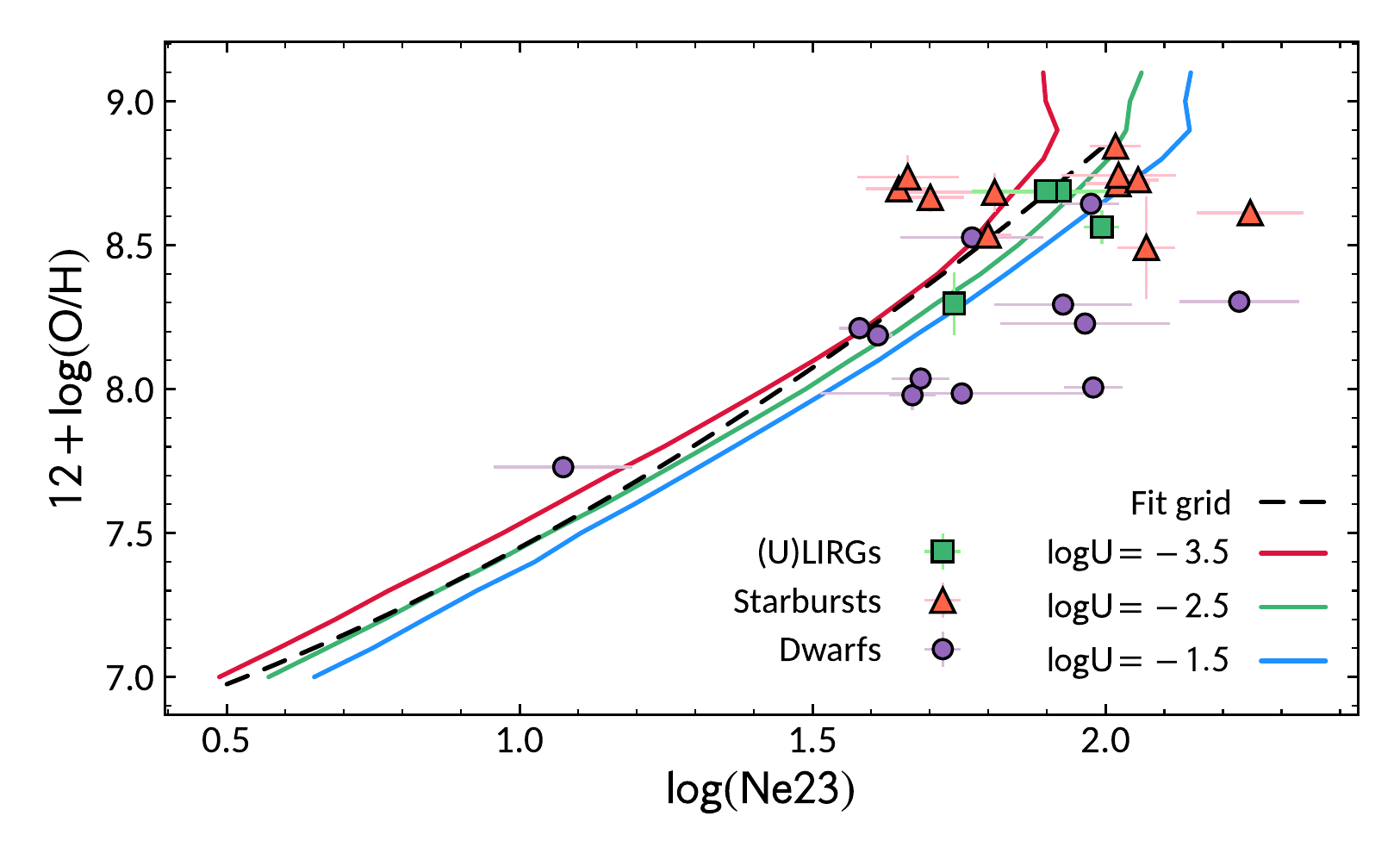}~
  \includegraphics[width=0.5\textwidth]{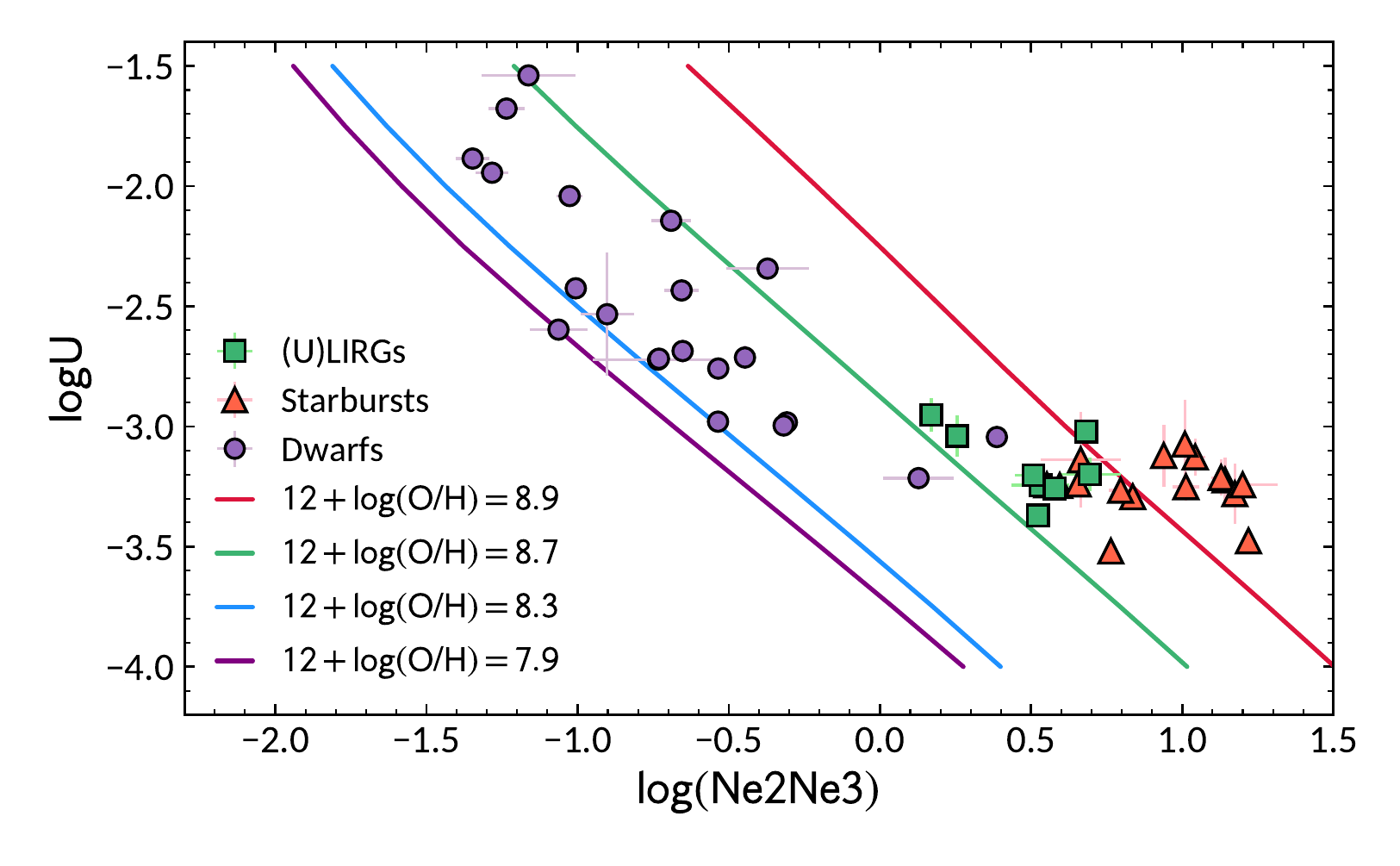}
  \caption{\textbf{Left:} relation between the Ne23 parameter and the total oxygen abundance for the subset of models with $\log(N/O) = -0.625\, \rm{dex}$. \textbf{Right:} relation between the Ne2Ne3 parameter and $\log U$. Note that the O/H and $\log U$ shown in these figures for the local star-forming galaxies (purple circles: dwarf galaxies; red triangles: nearby starbursts; green squares: local LIRGs) are derived from the optical emission lines using \textsc{HCm}, and compared with the IR parameters Ne23 and Ne2Ne3, respectively, measured in the IR spectra. A polynomial fit to the O/H values for the whole grid of models is also shown (black-dashed line on the left panel; see Eq.\,\ref{eq_ne23}).}\label{Ne23}%
\end{figure*}

\begin{figure*}
  \centering
  \includegraphics[width=0.5\textwidth]{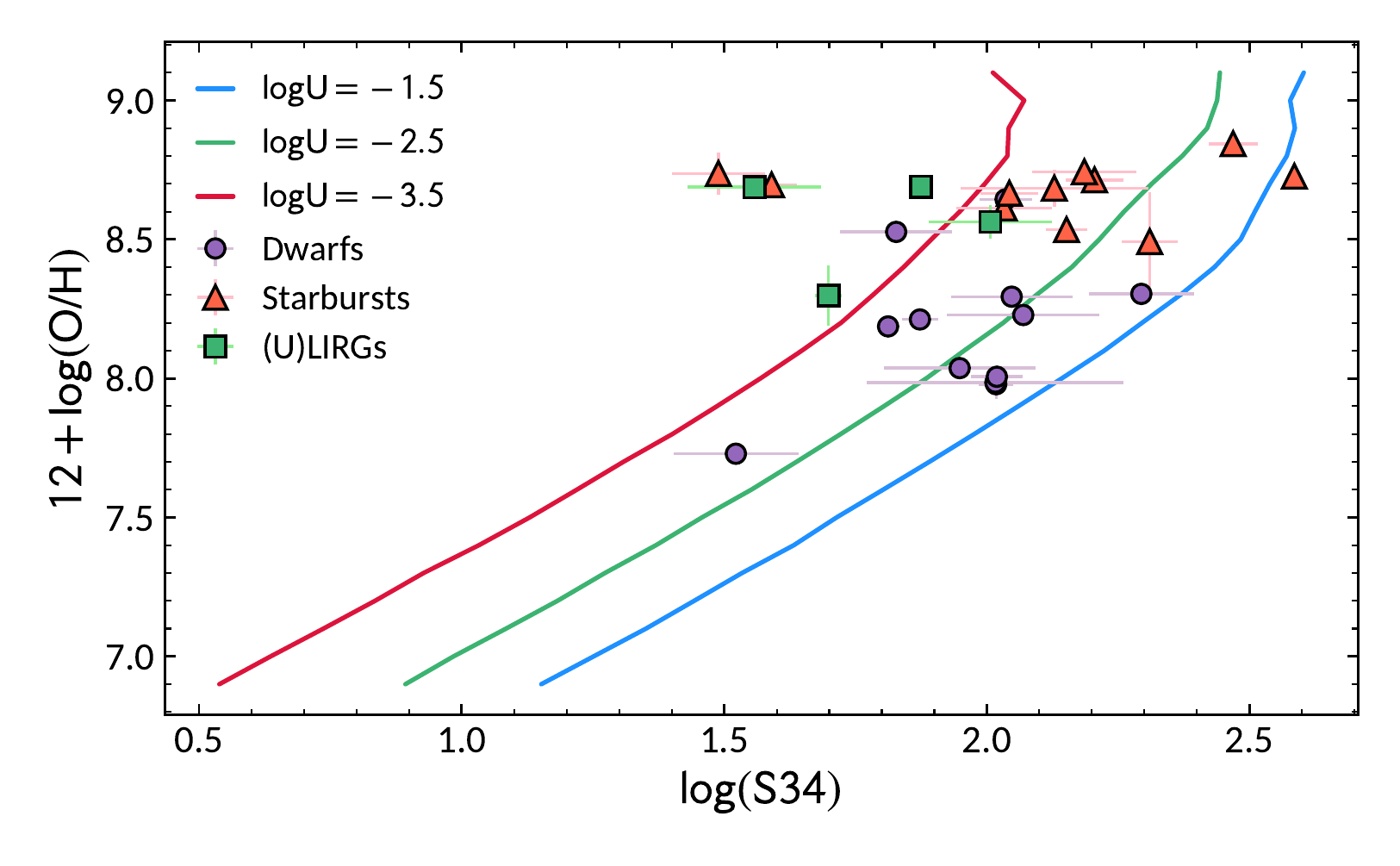}~
  \includegraphics[width=0.5\textwidth]{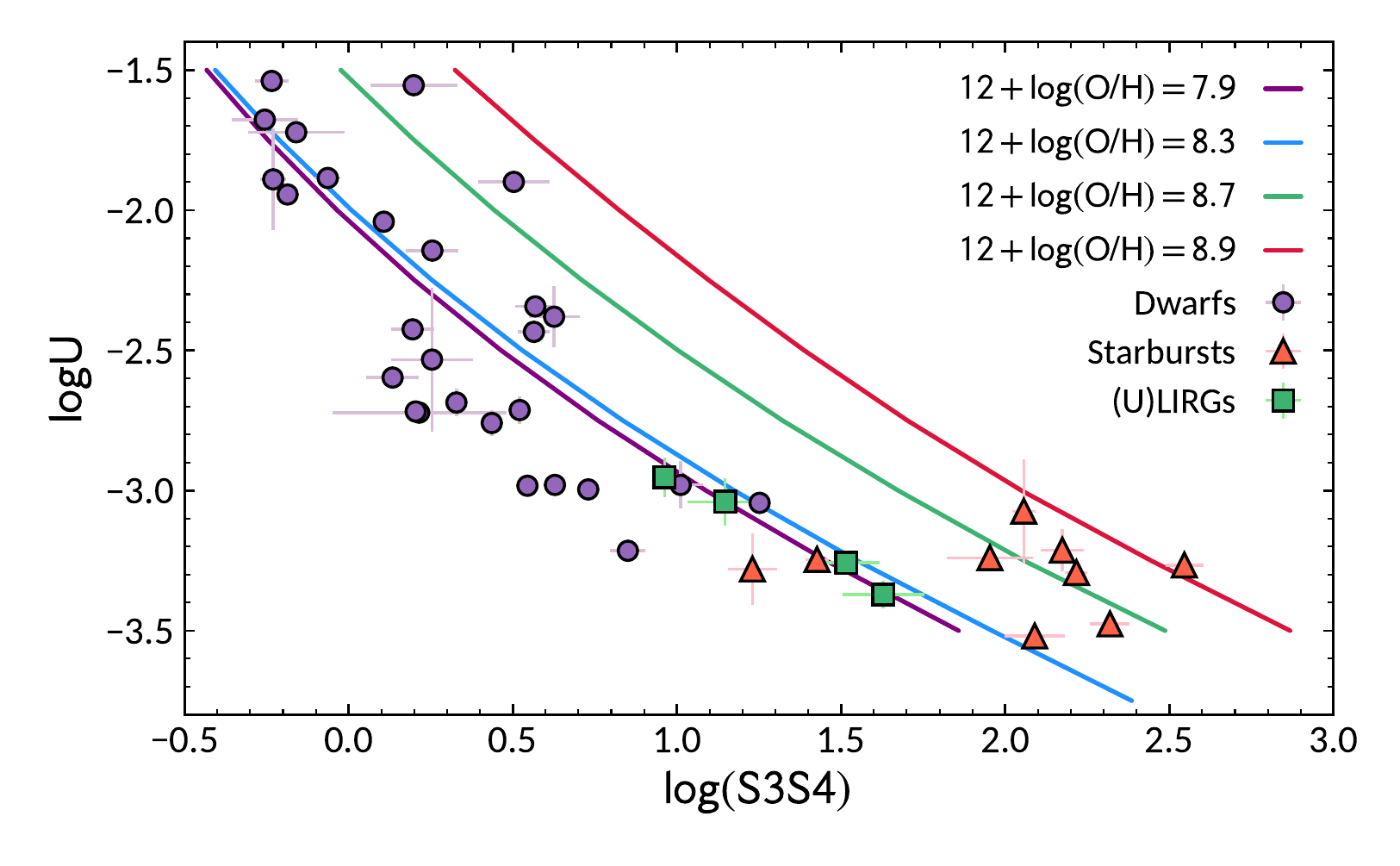}
  \caption{\textbf{Left:} relation between the S34 parameter, as calculated with \ion{H}{i} $12.37\, \rm{\micron}$, and the total oxygen abundance for the subset of models with $\log(N/O) = -0.625\, \rm{dex}$. \textbf{Right:} relation between the S3S4 parameter and $\log U$. The O/H and $\log U$ derived with \textsc{HCm} for the optical lines in a sample of local star-forming galaxies are compared with the IR parameters S34 and S3S4, respectively (purple circles: dwarf galaxies; red triangles: nearby starbursts; green squares: local LIRGs).}\label{S34}%
\end{figure*}

\begin{figure*}
  \centering
  \includegraphics[width=0.5\textwidth]{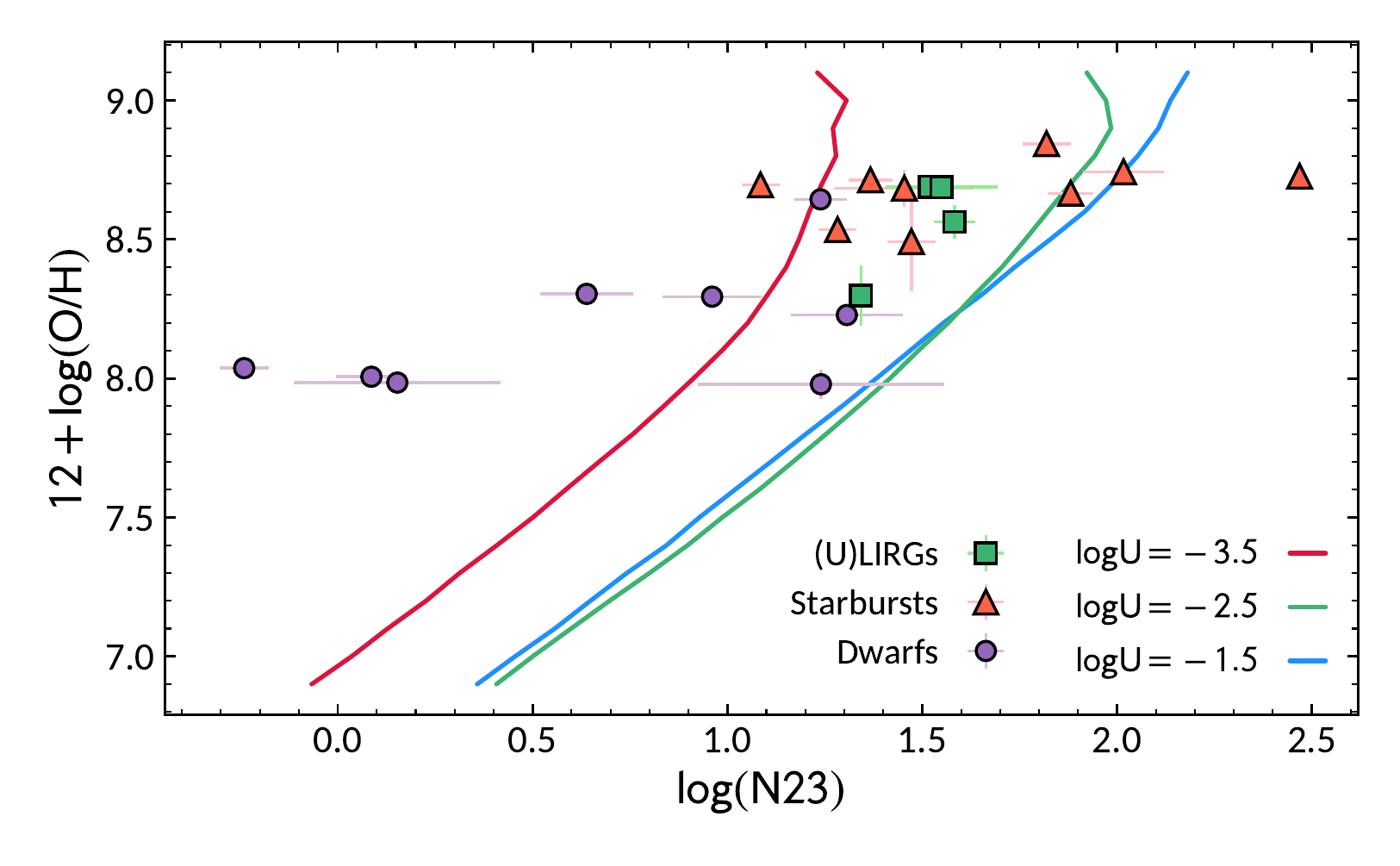}~
  \includegraphics[width=0.5\textwidth]{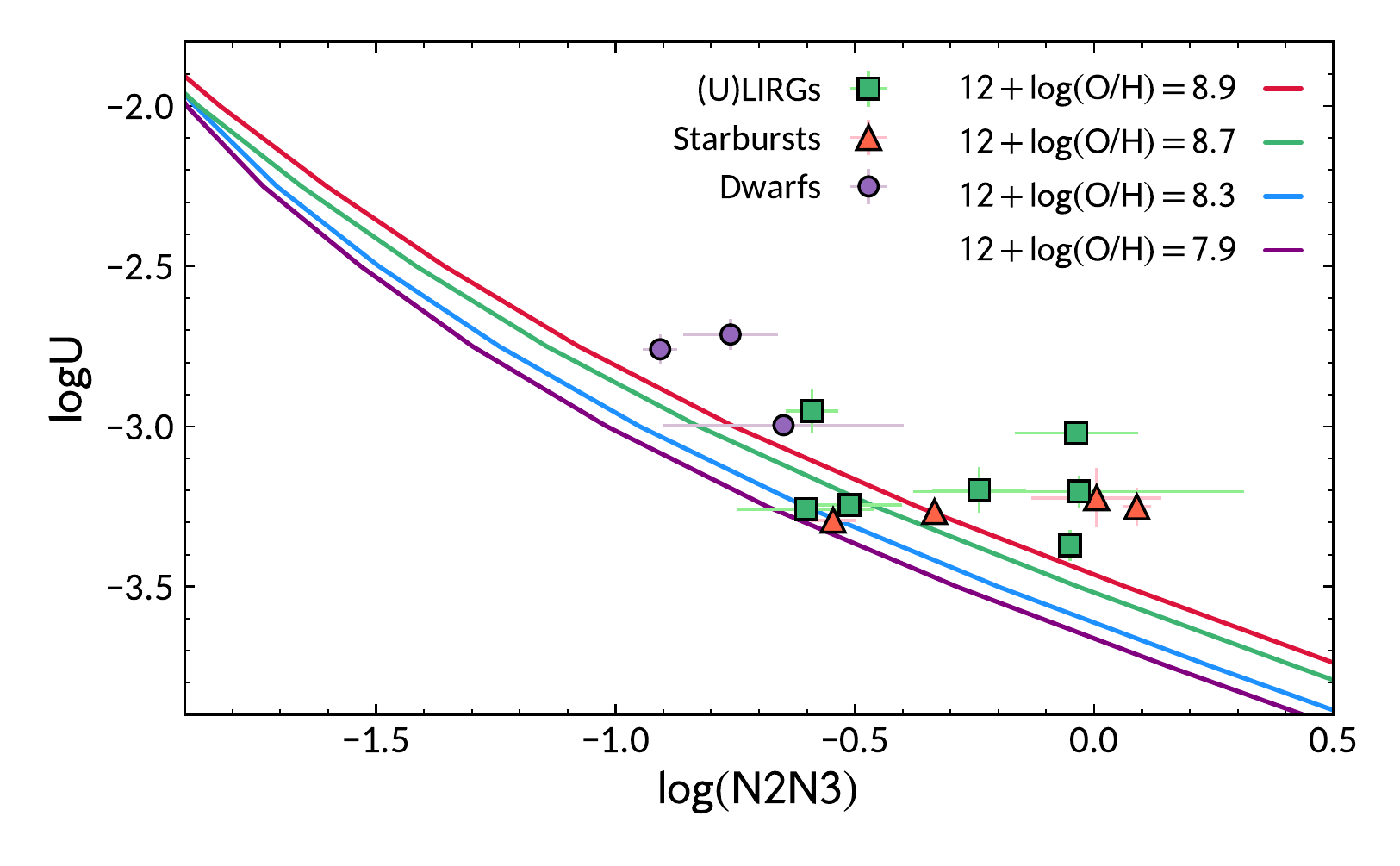}
  \caption{\textbf{Left:} relation between the N23 parameter, derived with \ion{H}{i} $12.37\, \rm{\micron}$, and the total oxygen abundance at a fixed N/O abundance of $-0.625\, \rm{dex}$. \textbf{Right:} relation between the N3N2 parameter and $\log U$. The O/H and $\log U$ derived with \textsc{HCm} for the optical lines in a sample of local star-forming galaxies are compared with the IR parameters N23 and N2N3, respectively (purple circles: dwarf galaxies; red triangles: nearby starbursts; green squares: local LIRGs).}\label{N23}%
\end{figure*}

\begin{figure*}
  \centering
  \includegraphics[width=0.5\textwidth]{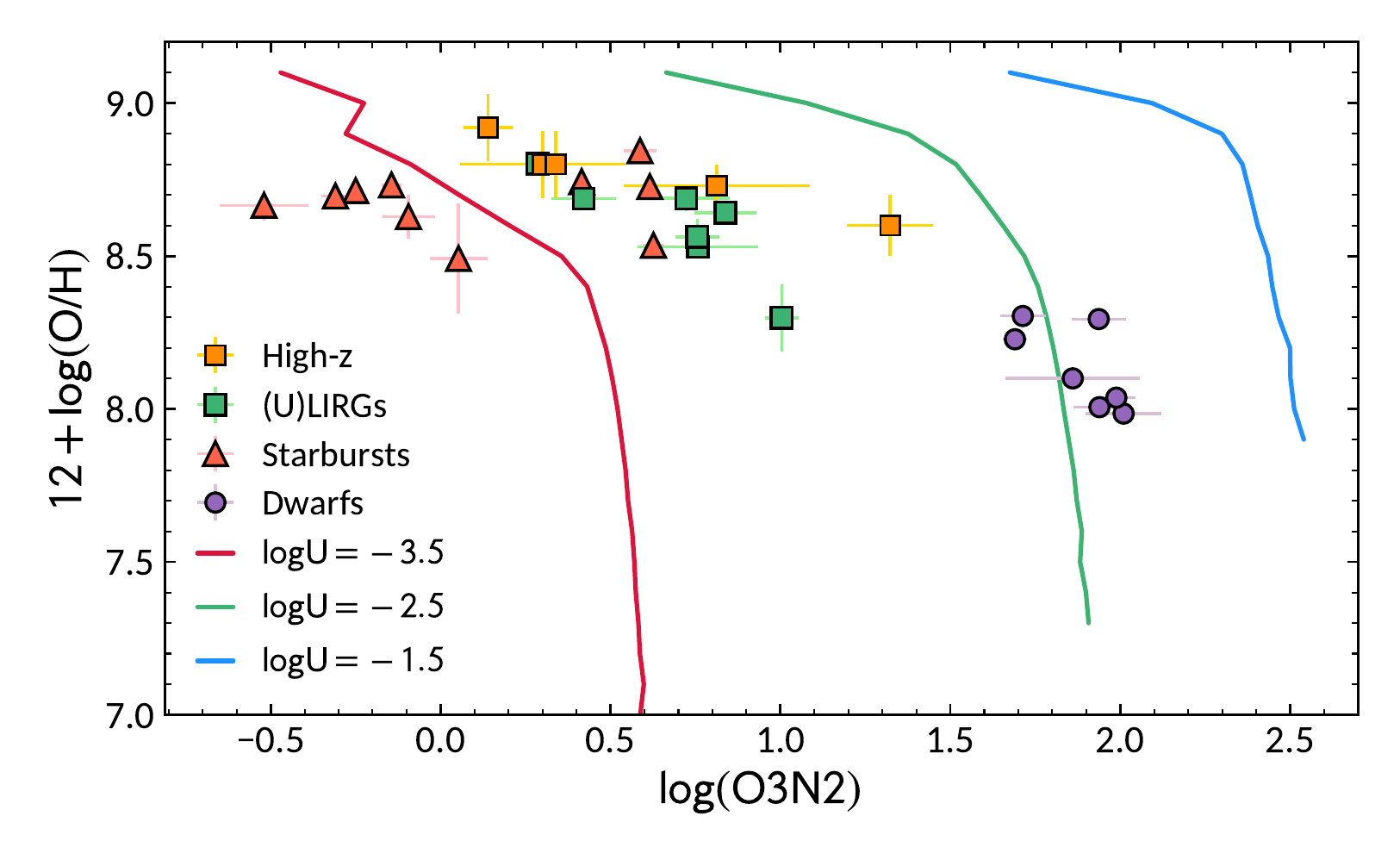}~
  \includegraphics[width=0.5\textwidth]{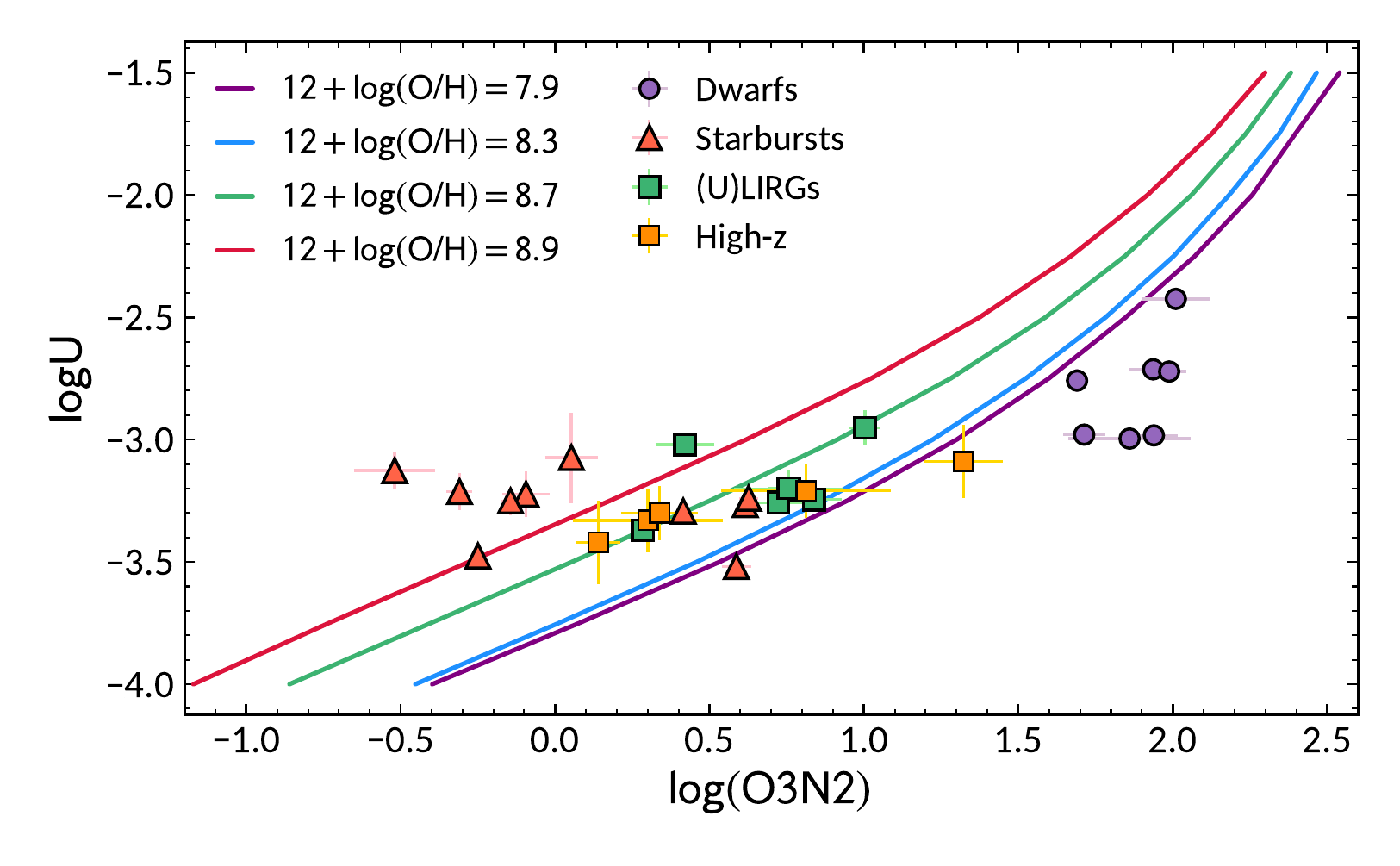}
  \caption{Relation between the infrared O3N2 parameter defined in this work and the total oxygen abundance (left panel) or $\log U$ (right panel) for the models at fixed $\log(N/O) = -0.625\, \rm{dex}$. The O/H and $\log U$ derived with \textsc{HCm} for the optical lines in our sample of local star-forming galaxies (purple circles: dwarf galaxies; red triangles: nearby starbursts; green squares: local LIRGs) and the sample of galaxies at high-$z$ (orange squares) compiled for this work (see Table\,\ref{tab_sample}).}\label{O3N2}%
\end{figure*}

In addition, as described in \cite{perezmontero2014}, in absence of any auroral emission line, such as [\ion{O}{iii}] $\lambda 4363$ for the optical range, the code has to make additional assumptions on the relation between O/H and $U$ in order to obtain results consistent with the $T_{\rm e}$-method. In particular, given that no auroral lines can be measured for our adapted version of \textsc{HCm-IR}, the code only considers in the calculations those models following the same relation between O/H and $U$ obtained empirically from the optical lines, described in \cite{perezmontero2014}, where objects with lower metallicity present a higher excitation. This also permits the use of emission-line ratios sensitive to excitation to be used as tracers of the total chemical content in the gas. 

Among the different emission-line ratios from the IR that the code uses are those depending on [\ion{Ne}{ii}] $12.8\, \rm{\micron}$ and [\ion{Ne}{iii}] $15.6\, \rm{\micron}$, such as the Ne23 parameter, defined as:
\begin{equation}
{\rm Ne}23 = \log \left( \frac{\rm I([\ion{Ne}{ii}]_{12.8 \mu{\rm m}}) + I([\ion{Ne}{iii}]_{15.6 \mu{\rm m}})}{\rm I(H_i)} \right)
\end{equation} 
where H$_{\rm i}$ represents any of the hydrogen recombination lines that can be measured in the mid-IR spectrum, specifically Br$\alpha$ $4.05\, \rm{\micron}$, Pf$\alpha$ $7.46\, \rm{\micron}$ and Hu$\alpha$ $12.37\, \rm{\micron}$ are used in the code. In the left panel of Fig.\,\ref{Ne23} we represent, for several model-sequences and the sample of objects, the relation between Ne23 and total oxygen abundance. As in Fig.\,\ref{n3o3}, the O/H values for the sample of galaxies have been obtained from the optical emission lines using \textsc{HCm}, which are compared with the Ne23 parameter derived from the IR nebular lines. This relation is very tight and presents a very high value for the correlation coefficient $\rho$ = 0.98, with very low dependence on $U$ and monotonically growing up to over-solar metallicities. We can also provide a polynomial fit to the whole grid of models (black-dashed line in Fig.\,\ref{Ne23}), giving as a result,
\begin{equation}\label{eq_ne23}
12+\log({\rm O/H}) = (6.65 \pm 0.02) + (0.50 \pm 0.03) \cdot x + (0.30 \pm 0.01) \cdot x^2 
\end{equation}
where $x$ = log(Ne23) is defined using the Hu$\alpha$ $12.37\, \rm{\micron}$ line. The standard deviation of the associated residuals is only $0.12\, \rm{dex}$. Another combination of these lines used by the code to estimate both O/H and $U$ through the space of models is Ne2Ne3, defined as:
\begin{equation}
{\rm Ne2Ne3} = \log \left( \frac{\rm I([\ion{Ne}{ii}]_{12.8 \mu{\rm m}})}{\rm I([\ion{Ne}{iii}]_{15.6 \mu{\rm m}})} \right) 
\end{equation}
that is represented in the right panel of Fig.\,\ref{Ne23}, also compared with the $U$ values derived from the optical lines for the observed galaxies. An equivalent expression for this ratio was calibrated to derive $U$ by \cite{thornley2000} and \cite{yeh2012}, although \cite{kewley2019} claim that it also depends on O/H for large values of $Z$.

For the sulphur lines, the code uses as an observable the S34 parameter, defined as:
\begin{equation}
{\rm S}34 = \log \left( \frac{\rm I([\ion{S}{iii}]_{18.7 \mu{\rm m}}) + I([\ion{S}{iii}]_{33.5 \mu{\rm m}}) + I([\ion{S}{iv}]_{10.5 \mu{\rm m}})}{\rm I(H_i)} \right)
\end{equation} 

We show in the left panel of Fig.\,\ref{S34} the relation of this parameter with O/H for different sequences of models at different values of $U$. This dependence on the ionisation parameter $U$ can be partially reduced by means of the S3S4 emission-line ratio, that we can define as:
\begin{equation}
{\rm S3S4} = \log \left( \frac{\rm I([\ion{S}{iii}]_{18.7 \mu{\rm m}}) + I([\ion{S}{iii}]_{33.5 \mu{\rm m}})}{\rm I([\ion{S}{iv}]_{10.5 \mu{\rm m}})} \right)
\end{equation}
whose relation with $U$ is also represented in the right panel of Fig.\,\ref{S34}. A similar emission line ratio, although without the [\ion{S}{iii}] line at $33.5\, \rm{\micron}$, was proposed by \cite{yeh2012} to trace the excitation in a way relatively independent of metallicity. Notice that the code is able to use these observable even if one of the two [\ion{S}{iii}] lines is absent, as the observables are defined as a function of the available observed information.

Finally, we also considered the code observables based on the nitrogen lines. This is the case of the N23 parameter, defined as:
\begin{equation}
{\rm N}23 = \log \left( \frac{\rm I([\ion{N}{ii}]_{122 \mu{\rm m}}) + I([\ion{N}{iii}]_{57 \mu{\rm m}})}{{\rm I(H_i)}} \right)
\end{equation} 
that is represented in the left panel of Fig.\,\ref{N23}. Its dependence on N/O is reduced if a previous determination of N/O, using the N3O3 is possible. Regarding its dependence on $U$ can be also minimised by means of the N2N3 parameter, that we can define as:
\begin{equation}
{\rm N2N3} = \log \left( \frac{\rm I([\ion{N}{ii}]_{122 \mu{\rm m}})}{\rm I([\ion{N}{iii}]_{57 \mu{\rm m}})} \right)
\end{equation} 

This ratio is also used by \cite{nagao2011} and \cite{kewley2019} as a tracer for $U$. These latter authors claim that it is almost insensitive to O/H, as can be seen also from Fig.\,\ref{N23} (right), although it can depend on gas pressure at high metallicities. We represent the relation of this parameter with $\log U$ in the right panel of Fig.\,\ref{N23}.

Another observable based on nitrogen lines that can be useful for high-redshift objects observed in the sub-mm regime is O3N2, defined as:
\begin{equation}\label{eq_o3n2}
{\rm O3N2} = \log \left( \frac{I([\ion{O}{iii}]_{88 \mu{\rm m}})}{\rm I([\ion{N}{ii}]_{122 \mu{\rm m}})} \right)
\end{equation} 

The relation between this ratio and the total oxygen abundance as derived both from models and from the optical lines is shown in Figure \ref{O3N2}. This relation is very similar to that observed between the O3N2 parameter defined using optical emission-lines \citep{pp04}, with an inverse relation with O/H for high metal contents, but a relatively flat behaviour for low values of O/H. At the same time, there is a strong dependence both on $U$ and N/O, that can be reduced under certain hypothesis on the relation of these with O/H. Due to the similar critical densities of [\ion{O}{iii}]$88\, \rm{\micron}$ ($510\, \rm{cm^{-3}}$) and [\ion{N}{ii}]$122\, \rm{\micron}$ ($310\, \rm{cm^{-3}}$), the ratio of these lines is insensitive to the gas density.

The code will provide a solution when at least one of the above described emission-line ratios is available, but the accuracy and precision of the derived abundance ratios and $\log U$ depend on the number of line ratios used in the computation. To analyse the accuracy of the code to recover these parameters when a reduced number of lines is available, we compared the known input model values with the code computations using different line combinations. The synthetic line fluxes used are based on the same emission-lines predicted by the models after applying a random 5\% perturbation to simulate the typical observational uncertainties in the line fluxes. In Table\,\ref{tab_comp1} we provide the mean offset and the standard deviation for the $\Delta$O/H, $\Delta$N/O, and $\Delta\log U$ residuals, obtained from the difference between the model input parameters and the estimated values using a restricted line set. This Table can be also used as a reference to know what sets of lines lead to a solution when the complete set of valid emission lines is not available.
\begin{table*}
\centering
\footnotesize
\caption{Mean offsets and standard deviation values for the $\Delta$O/H, $\Delta$N/O, and $\Delta\log U$ residuals, computed as the difference between the model input parameters and the resulting values derived with {\sc HCm-IR} when a restricted set of emission lines, also predicted from the same {\sc popstar} models used as input, is used for the abundance ratios and $\log U$ calculation.}\label{tab_comp1}
\begin{tabular}{lccccccc}
\bf Input lines & \bf Grid  & \multicolumn{2}{c}{\bf $\Delta$O/H} & \multicolumn{2}{c}{\bf $\Delta$N/O} & \multicolumn{2}{c}{\bf $\Delta$log U}  \\
   & & Mean & $\sigma$ & Mean & $\sigma$ & Mean & $\sigma$ \\
\hline\\[-0.2cm]
All lines & 2  &    +0.04 &  0.09   &   0.00  &  0.01   &  +0.02   &  0.07  \\
{[\ion{Ne}{ii}]$_{12.8}$}, [\ion{Ne}{iii}]$_{15.6}$, [\ion{S}{iii}]$_{18.7,33.5}$, [\ion{S}{iv}]$_{10.5}$, [\ion{O}{iii}]$_{52,88}$, [\ion{N}{ii}]$_{122}$, [\ion{N}{iii}]$_{57}$   & 2 &+0.03 & 0.13 & 0.00 & 0.01 & +0.02 & 0.07 \\ 
{[\ion{O}{iii}]$_{52,88}$}, [\ion{N}{ii}]$_{122}$, [\ion{N}{iii}]$_{57}$ & 2 & -0.01 & 0.18 & 0.00 & 0.01 & +0.02 & 0.15 \\ 
\ion{H}{i}, [\ion{S}{iii}]$_{18.7,33.5}$, [\ion{S}{iv}]$_{10.5}$, [\ion{Ne}{ii}]$_{12.8}$, [\ion{Ne}{iii}]$_{15.6}$  & 3 & +0.04 & 0.04 & -- & -- & -0.01 & 0.01 \\ 
{[\ion{S}{iii}]$_{18.7,33.5}$}, [\ion{S}{iv}]$_{10.5}$, [\ion{Ne}{ii}]$_{12.8}$, [\ion{Ne}{iii}]$_{15.6}$  & 3 & +0.04 & 0.04 & -- & -- & 0.00 & 0.01 \\ 
{[\ion{S}{iii}]$_{18.7,33.5}$}, [\ion{S}{iv}]$_{10.5}$  & 3 & +0.02& 0.02 & -- & -- & 0.00 & 0.01 \\ 
{[\ion{Ne}{ii}]$_{12.8}$}, [\ion{Ne}{iii}]$_{15.6}$  & 3 & +0.03 & 0.09 & -- & -- & 0.00 & 0.06 \\ 
{[\ion{N}{ii}]$_{122}$}, [\ion{N}{iii}]$_{57}$ & 3 & 0.00 & 0.11 & -- & -- & 0.00 & 0.06 \\[0.1cm]
\hline
\end{tabular}
\end{table*}

\begin{figure*}
  \centering
  \includegraphics[width=0.5\textwidth]{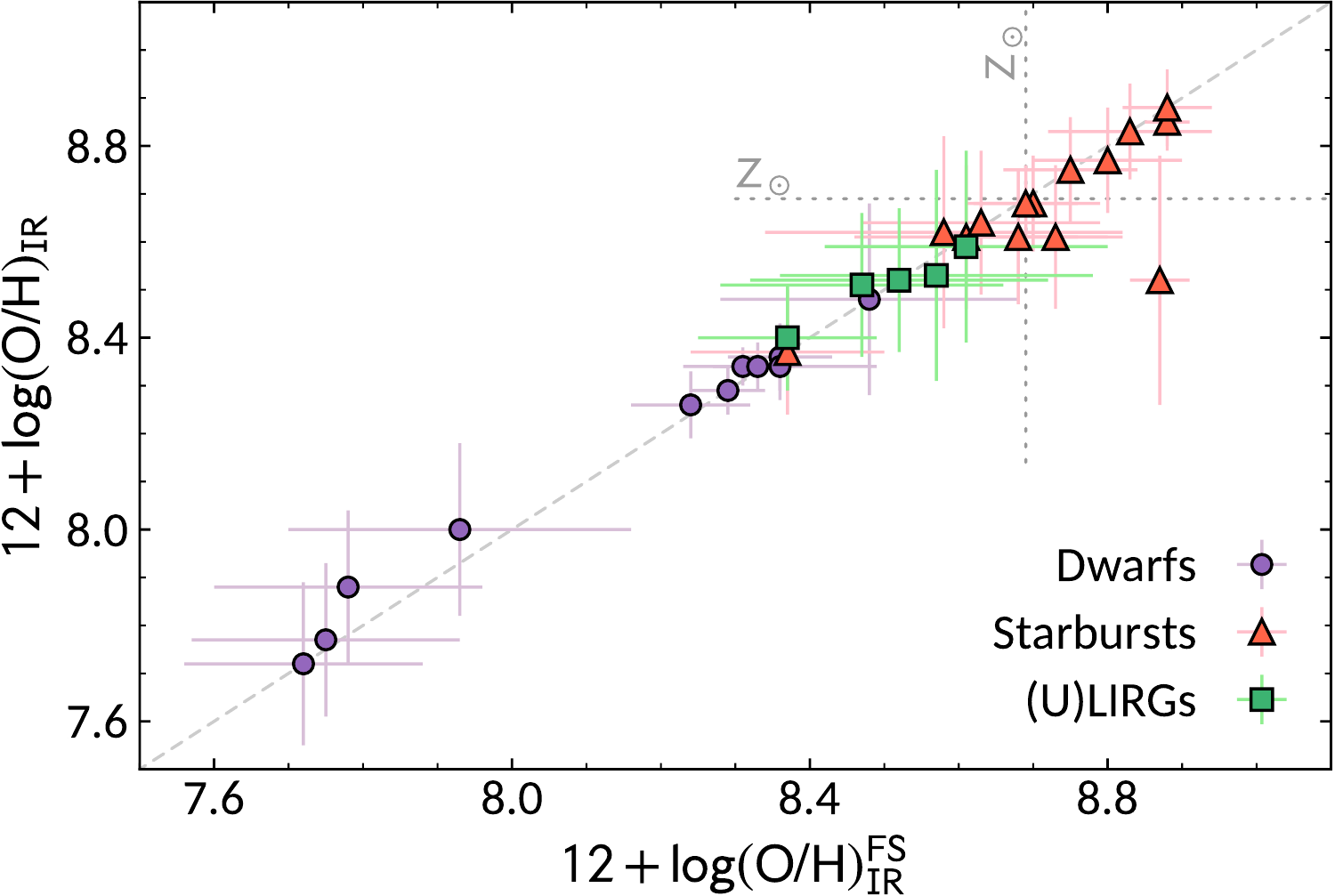}~
  \includegraphics[width=0.5\textwidth]{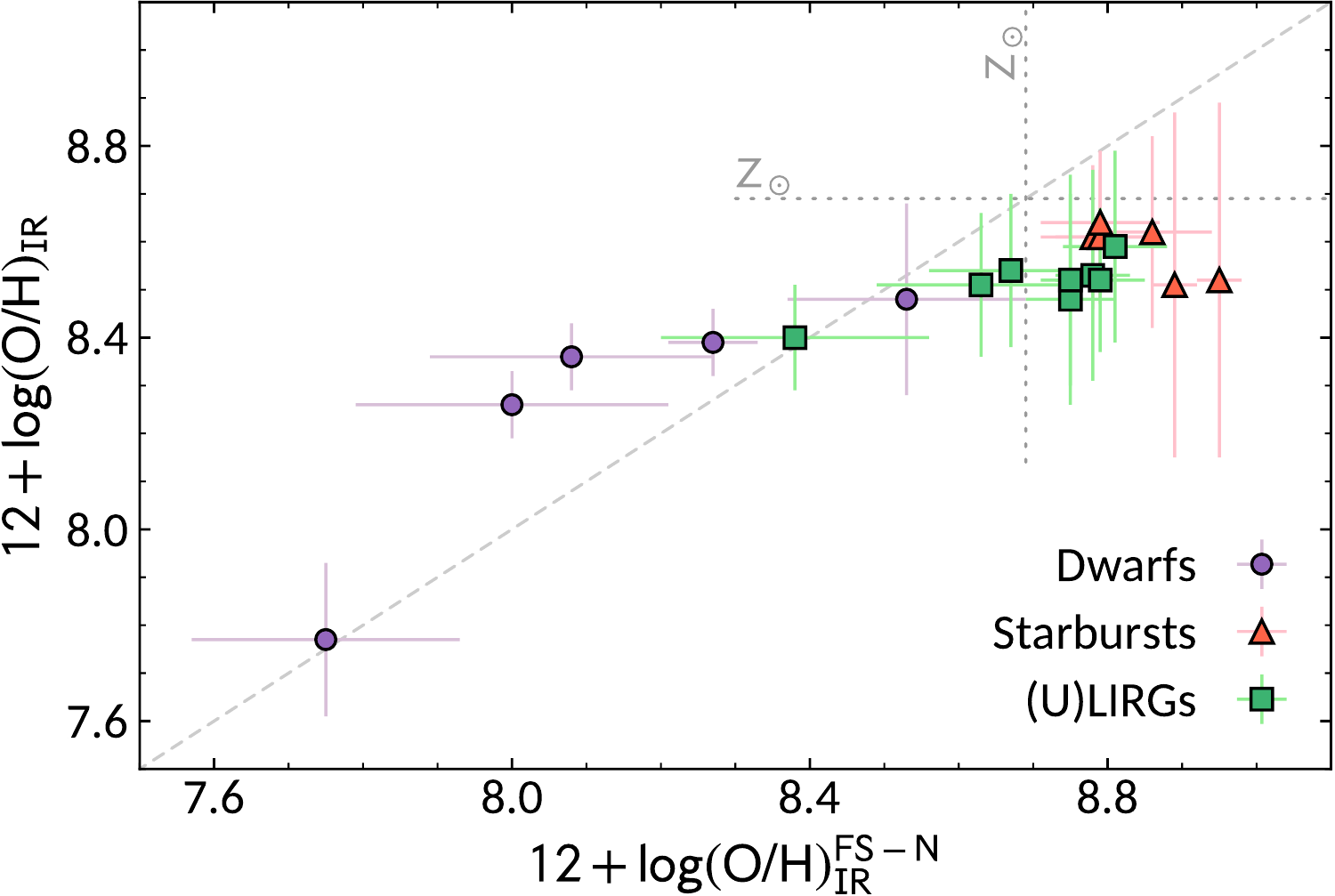}
  \caption{\textbf{Left:} comparison of the O/H abundances obtained with \textsc{HCm-IR} including the hydrogen recombination lines, $\log(O/H)_{IR}$, against the values obtained when only the IR fine-structure lines are used in the analysis, $\log(O/H)^{FS}_{IR}$. The sample includes dwarf galaxies (purple circles), nearby starbursts (red triangles), and local (U)LIRGs (green squares). \textbf{Right:} comparison of the O/H abundances obtained with \textsc{HCm-IR} including hydrogen and nitrogen lines against the values obtained with the IR fine-structure lines excluding the nitrogen lines, $\log(O/H)^{FS-N}_{IR}$. The sample includes dwarf galaxies (purple circles), nearby starbursts (red triangles), and local (U)LIRGs (green squares). In both panels the bisector of the diagram is indicated by the grey dashed line, while the solar metallicity values are marked in both axes by grey dotted lines.}\label{fig_noHN}
\end{figure*}


The results obtained for the different line combinations in Table\,\ref{tab_comp1} show minor offset deviations in the three parameters $\Delta$O/H, $\Delta$N/O, and $\Delta \log U$ ($\lesssim 0.04\, \rm{dex}$). This is comparable to the typical line flux errors ($\sim 10\%$) meaning that, provided a line ratio sensitive to the ionisation parameter such as [\ion{Ne}{iii}]$_{\rm 15.6 \mu m}$/[\ion{Ne}{ii}]$_{\rm 12.8 \mu m}$, [\ion{S}{iv}]$_{\rm 10.5 \mu m}$/[\ion{S}{iii}]$_{\rm 18.7,33.5 \mu m}$ or [\ion{N}{iii}]$_{\rm 57 \mu m}$/[\ion{N}{ii}]$_{\rm 122 \mu m}$, the code does not introduce any detectable systematics when the O/H and N/O relative abundances are derived using ionic species from different elements. The accuracy of the estimated abundances improves when the neon and/or the sulphur lines are available, as shown by the lower scatter measured in the corresponding line combinations in Table\,\ref{tab_comp1}. However, including the [\ion{O}{iii}]$_{\rm 52,88 \mu m}$ and [\ion{N}{iii}]$_{\rm 57 \mu m}$ lines is essential to provide an independent measurement of the N/O relative abundance, which is tightly constrained by this ratio given the similar ionisation potentials for the first ionisation stages of oxygen and nitrogen \citep[e.g.][]{pereira-santaella2017}.

\section{Results}\label{results}

In this section we use \textsc{HCm-IR} to derive chemical abundances based on fine-structure lines in the mid- to far-IR range for the three samples of star-forming galaxies described in Section\,\ref{data}. The values obtained are listed in Table\,\ref{tab_abund}. The systematics of the model-dependent abundances derived when the hydrogen recombination lines are not available are assessed by excluding the Hu$\alpha$ and Br$\alpha$ lines from our abundance calculations in Section\,\ref{noH}. Missing the nitrogen to oxygen line ratios necessary to estimate the relative N/O abundance will introduce an additional uncertainty, discussed in Section\,\ref{NO}. Finally, these results are compared with the abundance ratios and $\log U$ values obtained from the optical emission lines in Section\,\ref{IR_opt}, to quantify the effect of dust obscuration and temperature for the different samples of galaxies.

\begin{table*}
  \small
  \setlength{\tabcolsep}{2.pt}
	\centering
	\caption{Columns 1--3 show the name, chemical abundance, and N/O abundance ratio derived from the IR line fluxes using \textsc{HCm-IR} for the sample of star forming galaxies used in this work. These are compared with the abundance determinations derived from the optical line fluxes using \textsc{HCm} in column 4 and 5, and those reported in the literature in columns 6 favouring the direct method (DM) over the strong-line methods (SL) when available (references in column 7). Columns 8 and 9 include the abundance estimated obtained using the R and S strong-line calibrations in \citet{pilyugin2016}, respectively. The complete version of the table is published in the online version of this paper.}\label{tab_abund}
	\begin{threeparttable}[b]
	\begin{tabular}{lcccccccc}
		\bf Name & \bf 12+log(O/H)$_{\rm IR}$ & \bf log(N/O)$_{\rm IR}$ & \bf 12+log(O/H)$_{\rm HCm}$ & \bf \bf log(N/O)$_{\rm HCm}$ & \bf 12+log(O/H)$_{\rm lit}$ & \bf Refs. & \bf 12+log(O/H)$^{\rm R}_{\rm PG16}$ & \bf 12+log(O/H)$^{\rm S}_{\rm PG16}$ \\[-0.2cm]\\
    (1) & (2) & (3) & (4) & (5) & (6) & (7) & (8) & (9) \\\hline\\[-0.2cm]
Haro\,11       & $8.26 \pm 0.07$ & $-1.2 \pm 0.1$ & $8.23 \pm 0.02$ & $-0.73 \pm 0.08$ & $8.23 \pm 0.03$ & 1 & $8.45$ & $8.45$ \\
IRAS00397-1312 & $8.4 \pm 0.2$ & -- & $8.58 \pm 0.03$ & $-0.65 \pm 0.08$ & -- & & -- & $8.64$ \\
NGC\,253       & $8.5 \pm 0.4$ & $-0.3 \pm 0.3$ & $8.63 \pm 0.07$ & $-0.5 \pm 0.1$ & $8.62 \pm 0.06$ & 2 & $8.77$ & $8.76$ \\
HS0052+2536    & $8.1 \pm 0.2$ & -- & -- & -- & $8.1 \pm 0.2$ & 1 & -- & -- \\
UM311          & $8.1 \pm 0.2$ & -- & $8.485 \pm 0.006$ & $-1.21 \pm 0.03$ & $8.37 \pm 0.05$ & 1 & $8.16$ & $8.17$ \\
NGC\,625       & $8.0 \pm 0.2$ & -- & $8.21 \pm 0.01$ & $-1.26 \pm 0.03$ & $8.18 \pm 0.03$ & 1 & $8.11$ & $8.11$ \\
NGC\,891       & $8.85 \pm 0.06$ & -- & $8.68 \pm 0.07$ & $-0.85 \pm 0.09$ & -- & & -- & $8.48$ \\
NGC\,1140      & $8.34 \pm 0.04$ & -- & $8.304 \pm 0.006$ & $-1.19 \pm 0.05$ & $8.27 \pm 0.08$ & 1 & $8.24$ & $8.27$ \\
NGC\,1222      & $8.4 \pm 0.1$ & -- & -- & -- & $8.57 \pm 0.09$ & 3 & -- & -- \\
SBS\,0335-052  & $7.8 \pm 0.2$ & -- & $7.34 \pm 0.02$ & $-1.34 \pm 0.05$ & $7.29 \pm 0.01$ & 1 & $7.27$ & $7.30$ \\[0.1cm]
		\hline
	\end{tabular}
	\begin{tablenotes}
	\item (1) \citealt{madden2013} (DM); (2) \citealt{pilyugin2014} (SL); (3) \citealt{petrosian1993} (SL).
    \end{tablenotes}
  \end{threeparttable}
\end{table*}

\begin{table}
  \centering
  \footnotesize
  \setlength{\tabcolsep}{1.8pt}
  \caption{Mean offsets and standard deviation values for the samples of dwarf galaxies, starbursts, and (U)LIRGs derived from the comparison ($\Delta$O/H) between the IR abundances obtained with \textsc{HCm-IR} and the different optical-based methods, namely \textsc{HCm}, the direct method (DM), and the \textit{R} and \textit{S} strong-line calibrations from \citet{pilyugin2016}. Results in the last three columns correspond to the total sample of galaxies.}\label{tab_comp2}
\begin{tabular}{lcccccccccccc}
\bf $\Delta$O/H & \multicolumn{3}{c}{\bf Dwarfs} & \multicolumn{3}{c}{\bf Starbursts} & \multicolumn{3}{c}{\bf (U)LIRGs} & \multicolumn{3}{c}{\bf Total} \\
   & Mean & $\sigma$ & N & Mean & $\sigma$ & N & Mean & $\sigma$ & N & Mean & $\sigma$ & N \\
\hline\\[-0.2cm]
 \textsc{HCm} & -0.08 & 0.24 & 25 & -0.04 & 0.16 & 16 & +0.12 & 0.12 &  8 & -0.03 & 0.21 & 49 \\
 DM           & -0.08 & 0.23 & 28 &    -- &   -- & -- &    -- &   -- & -- &    -- &   -- & -- \\ 
 PG16-R       & -0.03 & 0.22 & 22 & -0.11 & 0.22 & 11 & +0.13 &   -- &  3 & -0.04 & 0.22 & 36 \\ 
 PG16-S       & -0.05 & 0.24 & 23 & -0.12 & 0.21 & 14 & +0.13 & 0.05 &  8 & -0.04 & 0.23 & 45 \\[0.1cm]
\hline
\end{tabular}
\end{table}

\subsection{Dependence on hydrogen recombination lines}\label{noH}
The weak dependency of IR nebular line emissivities to the gas temperature (Fig.\,\ref{fig_emiss}) is one of the main arguments in favour of IR- over optical-based abundances tracers, but at the same time limits our ability to perform the same kind of analysis done in \citet{perezmontero2014}, where model-based abundances computed with \textsc{HCm} were compared with those obtained using the direct method for the same set of optical emission lines. Comparing optical direct method abundances with IR-based determinations is expected to introduce additional differences due to dust obscuration and temperature effects, the latter associated with the spatially unresolved temperature distribution and possible contributions from cold ionised gas components. Both obscuration and temperature affect the optical line fluxes and therefore the direct method estimates (see Section\,\ref{IR_opt}).

Nevertheless, model-independent IR-based abundances can be derived when a hydrogen recombination line is measured. The main hydrogen recombination lines in the mid- to far-IR range are Br$\alpha$ at $4.05\, \rm{\micron}$, Pf$\alpha$ at $7.46\, \rm{\micron}$, and Hu$\alpha$ at $12.37\, \rm{\micron}$. Br$\alpha$ is the brightest transition with $\sim 1/12$th of the H$\beta$ intensity (Case B, $T_{\rm e} = 10^4\, \rm{K}$, $n_{\rm e} = 100\, \rm{cm^{-3}}$; \citealt{osterbrock2006,luridiana2015}), however this line is out of the spectral range accessible to \textit{Spitzer} in nearby galaxies, and thus Br$\alpha$ fluxes are only available for 3 sources in our sample observed by \textit{AKARI}. Pf$\alpha$ ($\sim 1/39$th of H$\beta$) was observed by the IRS low-resolution module in \textit{Spitzer}, unfortunately the line sits on the ridge of a strong PAH feature at $7.7\, \rm{\micron}$ and a spectral resolution of $R \sim 50$ is not enough to deblend the line contribution from the PAH. Thus, the most observed recombination line in our sample is Hu$\alpha$, which is a relatively weak transition ($1/102$th of H$\beta$), although it was detected in 29 out of 72 galaxies in our sample ($40\%$).

To test the robustness of model-based abundances with \textsc{HCm-IR}, bearing in mind that hydrogen recombination lines are faint and hardly detected in the mid-IR range, we computed the IR abundances for our sample of galaxies including only the fine-structure lines. The comparison between the abundances obtained including the IR hydrogen recombination lines, $(O/H)_{IR}$, and those based only on the fine-structure lines, $(O/H)^{FS}_{IR}$, for 26 galaxies are shown in Fig.\,\ref{fig_noHN} (left panel). The overall agreement is excellent, with a median difference of $0.00$, a standard deviation of $0.03\, \rm{dex}$, and only one galaxy showing $0.35\, \rm{dex}$ lower abundances when Hu$\alpha$ was not included. Note that the $(O/H)_{IR}$ values obtained using \textsc{HCm-IR} are virtually independent of the photo-ionisation models adopted when an hydrogen recombination line is measured, since the abundances of the different ionic species relative to H are tightly constrained in that case \citep[e.g.][]{bernard-salas2001,wu2008,bernard-salas2009}. Thus, Fig.\,\ref{fig_noHN} proves that metallicity determinations based only on the mid-IR fine-structure lines are robust even in the absence of hydrogen recombination lines.

\subsection{Dependence on the N/O abundance}\label{NO}
Measuring the [\ion{O}{iii}]$_{\rm 52,88 \mu m}$ and [\ion{N}{iii}]$_{\rm 57 \mu m}$ lines is required to fix the relative N/O abundance in the photo-ionisation models. Otherwise, the code assumes a certain O/H--N/O relation (see Section\,\ref{OH-derivation}). To test the robustness of our method we removed both the nitrogen and the hydrogen recombination lines from the database and run the code to determine a new set of abundances, $(O/H)^{FS-N}_{IR}$. In the right panel of Fig.\,\ref{fig_noHN} we compare the values obtained for the 16 galaxies in our sample with a detection of [\ion{N}{iii}]$_{\rm 57 \mu m}$ and at least one of the [\ion{O}{iii}]$_{\rm 52,88 \mu m}$ lines. Both estimates are in agreement with a median difference of $0.18\, \rm{dex}$ in the abundances. For the dataset used in this work, the estimates missing the nitrogen and hydrogen lines provide a broader range in metallicity. That is, assuming a local O/H--N/O relation tends to underestimate the dwarf galaxy metallicities by $\sim 0.3$--$0.4\, \rm{dex}$ except for one object, while those of solar-like starburst galaxies are overestimated by a similar factor. This warns on the assumption of a solar-like N/O ratio when chemical abundances are obtained using nitrogen-based tracers, e.g. as in the case of high-z galaxies, since deviations from the solar values may likely introduce large systematics in the estimates. A similar situation is expected for carbon-based tracers.

\begin{table*}
\centering
\footnotesize
\setlength{\tabcolsep}{2.pt}
\caption{Average offsets and standard deviation obtained for each galaxy subsample of size N where the chemical abundances could be compared ($\Delta$O/H) between the various optical-based determinations and the \textsc{HCm-IR} estimates. The optical methods used are \textsc{HCm} from \citet{perezmontero2014}, the direct method (DM) values provided by \citet{madden2013}, and the strong-line calibrations from \citet{pilyugin2016}, namely the \textit{R} and \textit{S} calibrations. Note that the differences reported for the DM are estimated using only the sample of dwarfs galaxies, where DM abundances are available \citep{madden2013}.}\label{tab_comp3}
\begin{tabular}{lcccccccccccc}
\bf $\Delta$O/H & \multicolumn{3}{c}{\bf HCm} & \multicolumn{3}{c}{\bf DM} & \multicolumn{3}{c}{\bf PG16-R} & \multicolumn{3}{c}{\bf PG-16-S} \\
Input lines in \textsc{HCm-IR} & Mean & $\sigma$ & N & Mean & $\sigma$ & N & Mean & $\sigma$ & N & Mean & $\sigma$ & N \\
\hline\\[-0.2cm]
All lines & +0.02 & 0.10 & 5 & +0.02 & -- & 2 & +0.05 & 0.14 & 4 & +0.09 & 0.11 & 5 \\
{[\ion{Ne}{ii}]$_{12.8}$}, [\ion{Ne}{iii}]$_{15.6}$, [\ion{S}{iii}]$_{18.7,33.5}$, [\ion{S}{iv}]$_{10.5}$, [\ion{O}{iii}]$_{52,88}$, [\ion{N}{ii}]$_{122}$, [\ion{N}{iii}]$_{57}$ & +0.06 & 0.14 & 6 & +0.01 & 0.10 & 6 & +0.07 & 0.13 & 5 & +0.10 & 0.10 & 6 \\
{[\ion{O}{iii}]$_{52,88}$}, [\ion{N}{ii}]$_{122}$, [\ion{N}{iii}]$_{57}$ & -0.07 & 0.19 & 12 & +0.05 & 0.17 & 7 & +0.11 & 0.13 & 7 & +0.12 & 0.09 & 12 \\
\ion{H}{i}, [\ion{S}{iii}]$_{18.7,33.5}$, [\ion{S}{iv}]$_{10.5}$, [\ion{Ne}{ii}]$_{12.8}$, [\ion{Ne}{iii}]$_{15.6}$ & +0.02 & 0.15 & 21 & +0.08 & 0.14 & 11 & 0.00 & 0.17 & 18 & 0.00 & 0.17 & 20 \\
{[\ion{S}{iii}]$_{18.7,33.5}$}, [\ion{S}{iv}]$_{10.5}$, [\ion{Ne}{ii}]$_{12.8}$, [\ion{Ne}{iii}]$_{15.6}$ & +0.02 & 0.21 & 32 & +0.03 & 0.24 & 22 & -0.02 & 0.21 & 26 & -0.03 & 0.23 & 30 \\
{[\ion{S}{iii}]$_{18.7,33.5}$}, [\ion{S}{iv}]$_{10.5}$ & +0.09 & 0.24 & 37 & +0.11 & 0.27 & 27 & -0.02 & 0.21 & 30 & -0.03 & 0.22 & 34 \\
{[\ion{Ne}{ii}]$_{12.8}$}, [\ion{Ne}{iii}]$_{15.6}$ & -0.17 & 0.18 & 43 & -0.18 & 0.25 & 22 & -0.04 & 0.23 & 31 & -0.04 & 0.23 & 40 \\
{[\ion{N}{ii}]$_{122}$}, [\ion{N}{iii}]$_{57}$ & -0.17 & 0.12 & 12 & -0.14 & -- & 2 & +0.11 & 0.13 & 7 & +0.12 & 0.09 & 12 \\[0.1cm]
\hline
\end{tabular}
\end{table*}

\subsection{IR vs. optical abundances}\label{IR_opt}
To further assess the validity of the abundances derived with \textsc{HCm-IR} we have compared our results with those obtained from the optical lines using the direct method and the \textit{R} and \textit{S} strong-line calibrations from \citet{pilyugin2016}. The direct method relies on the temperature determination from the auroral lines and therefore provides the most reliable values in the optical range, while the two strong-line calibrations adopted are suitable for a wide range of metallicities and are consistent with the direct method estimates within an average offset of $\lesssim 0.05\, \rm{dex}$ \citep{pilyugin2016}. We note that the regions where the IR and the optical nebular lines are formed in galaxies do not generally match, since the IR emission traces also the obscured gas hidden by dust in the optical, while the optical line emission would be dominated by the hotter gas regions. Thus, an intrinsic scatter is always expected when optical and IR observables are compared, independently of the method adopted to derive the abundances.

In Fig.\,\ref{fig_IRopt} and Tables\,\ref{tab_abund}, \ref{tab_comp2} and \ref{tab_comp3} we compare the abundances obtained with \textsc{HCm-IR} with those derived from the optical lines using the direct method and the \textit{R} and \textit{S} strong-line calibrations. We note that direct method abundances were only available for the sample of dwarf galaxies \citep{madden2013}, where the [\ion{O}{iii}] $\lambda 4363$ auroral line is detected in sufficient numbers, and therefore the comparison between \textsc{HCm-IR} and the direct method (upper panels in Fig.\,\ref{fig_IRopt}) is limited to the dwarf galaxy sample. This should also mitigate the possible scatter due to dust obscuration, which is expected to be stronger in more metallic and dustier galaxies. For the strong-line calibrations (middle panels in Fig.\,\ref{fig_IRopt}), the comparison with \textsc{HCm-IR} could be extended to the starburst and (U)LIRGs samples. To derive the optical-based abundances with \textsc{HCm} and the strong-line calibrations we used the same line fluxes adopted by \citet{madden2013} for the direct method determinations.

\begin{figure*}[h!!!!]
  \centering
  \includegraphics[width=0.49\textwidth]{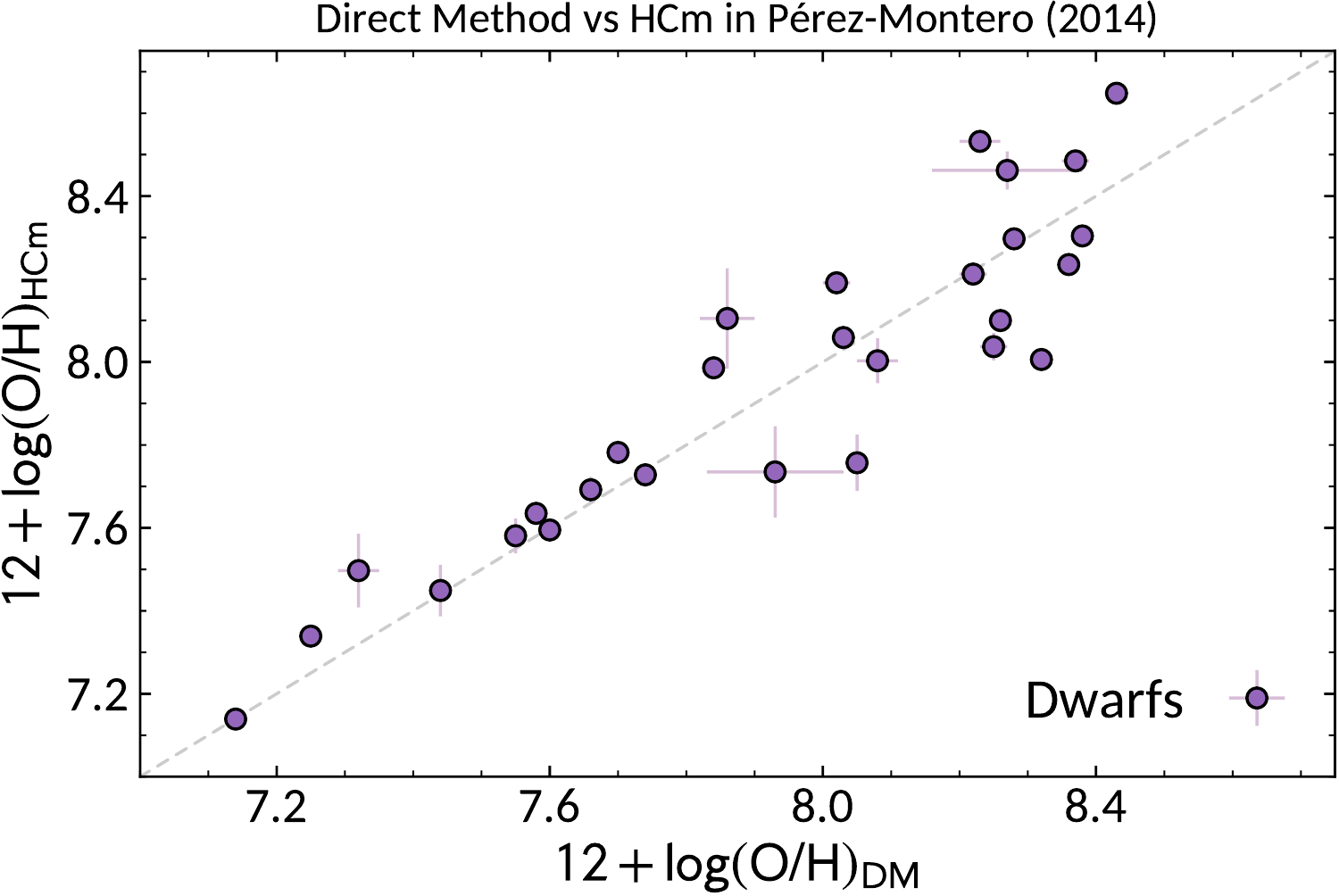}~
  \includegraphics[width=0.49\textwidth]{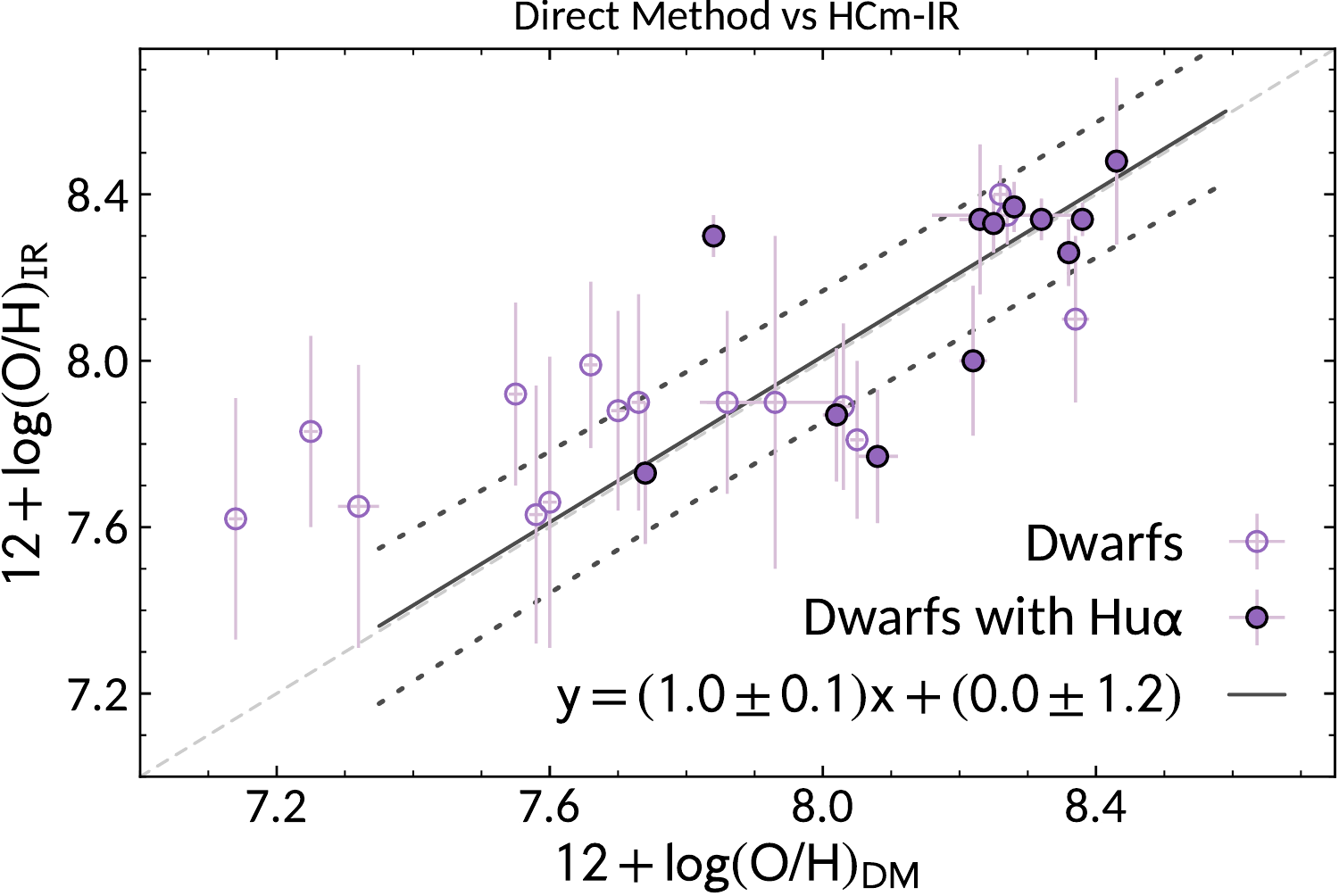}\\[0.2cm]
  \includegraphics[width=0.49\textwidth]{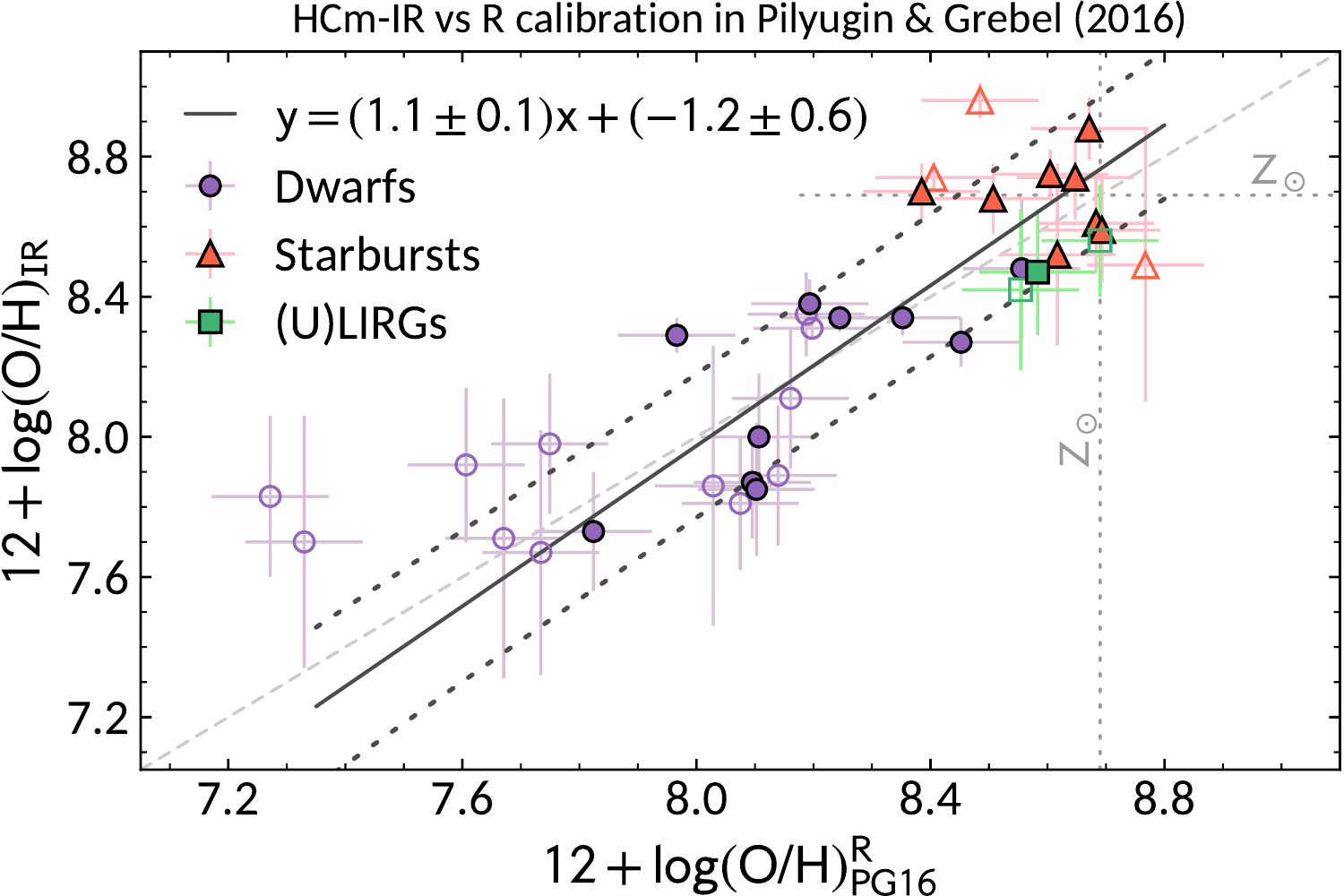}~
  \includegraphics[width=0.49\textwidth]{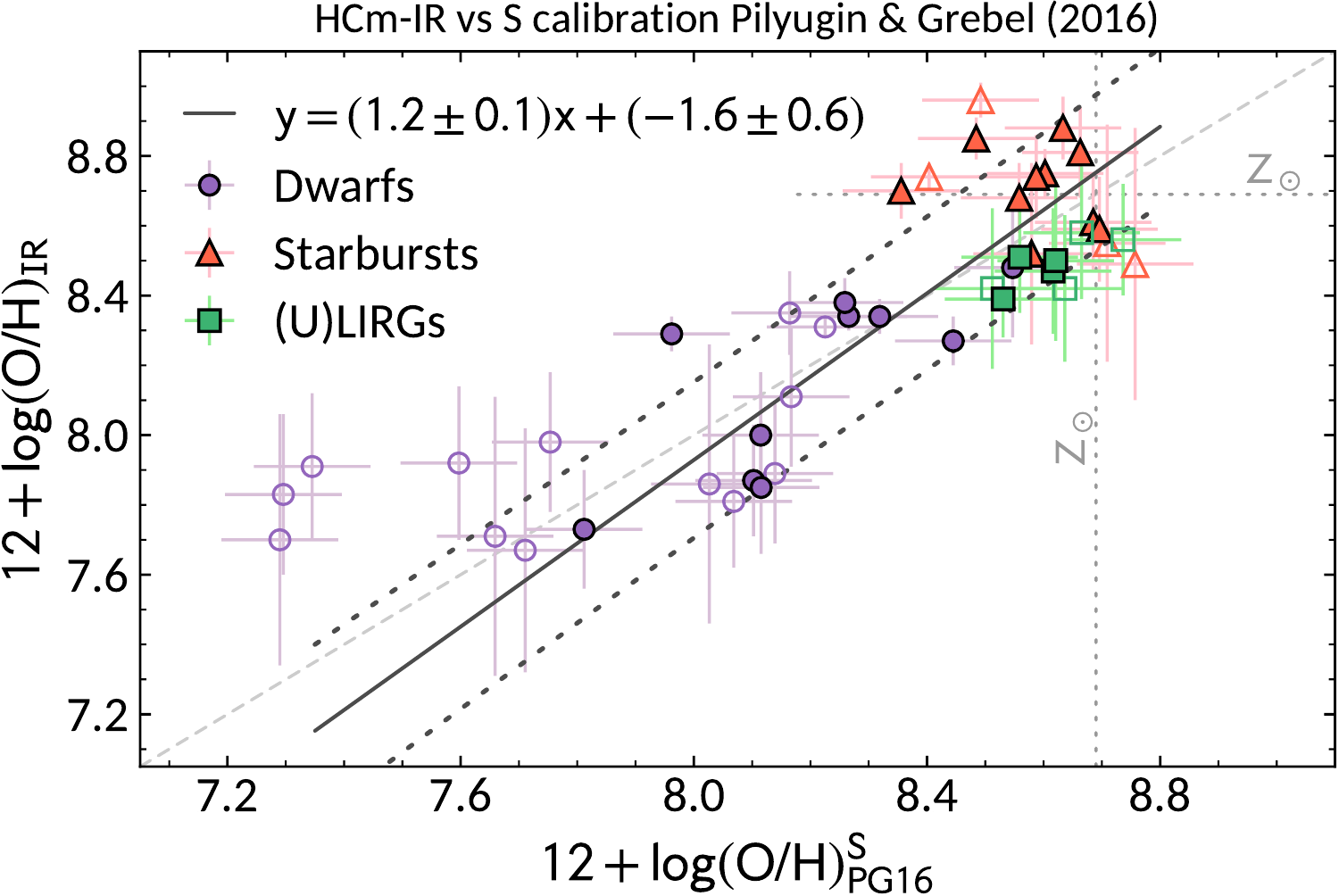}\\[0.2cm]
  \includegraphics[width=0.49\textwidth]{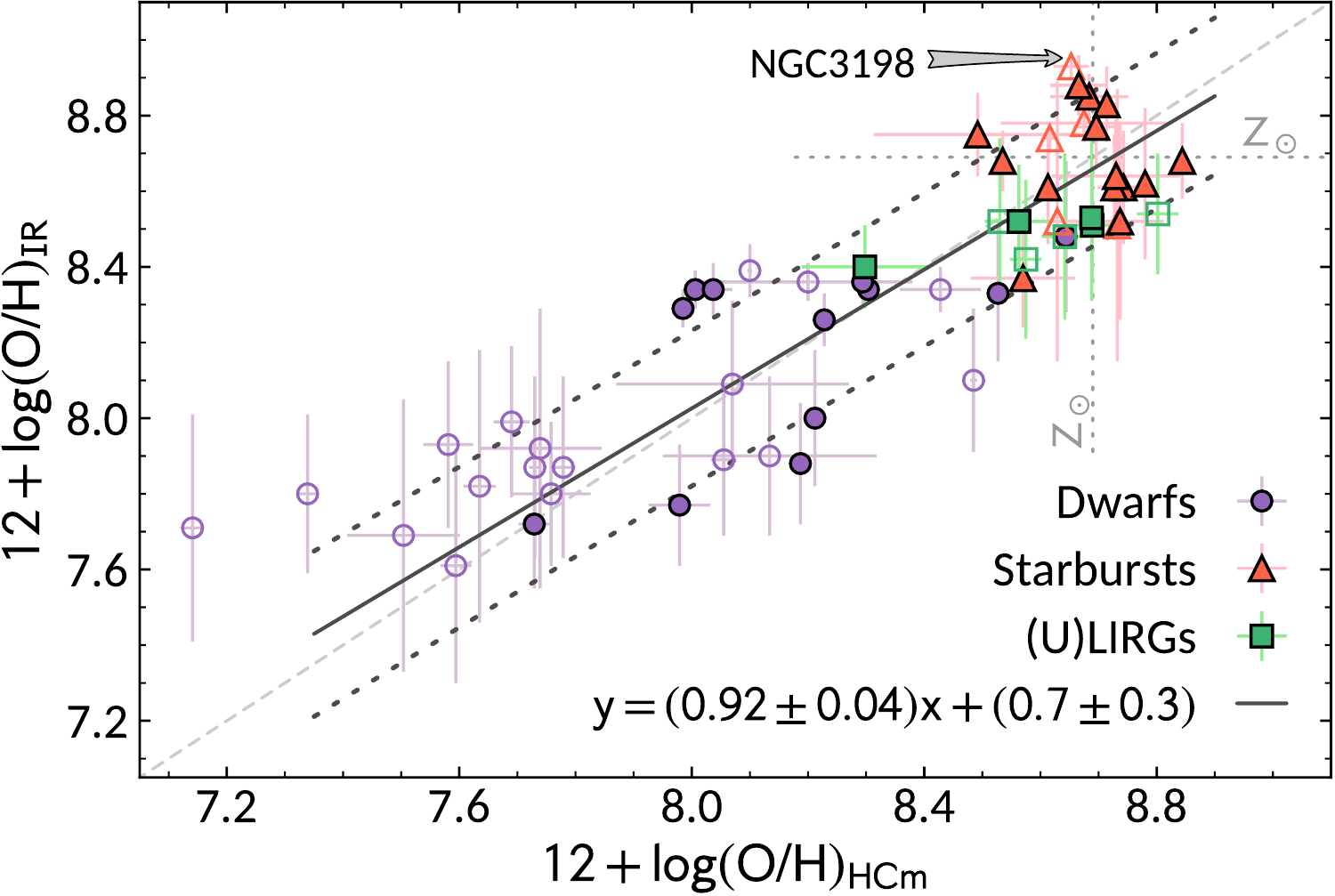}~
  \includegraphics[width=0.49\textwidth]{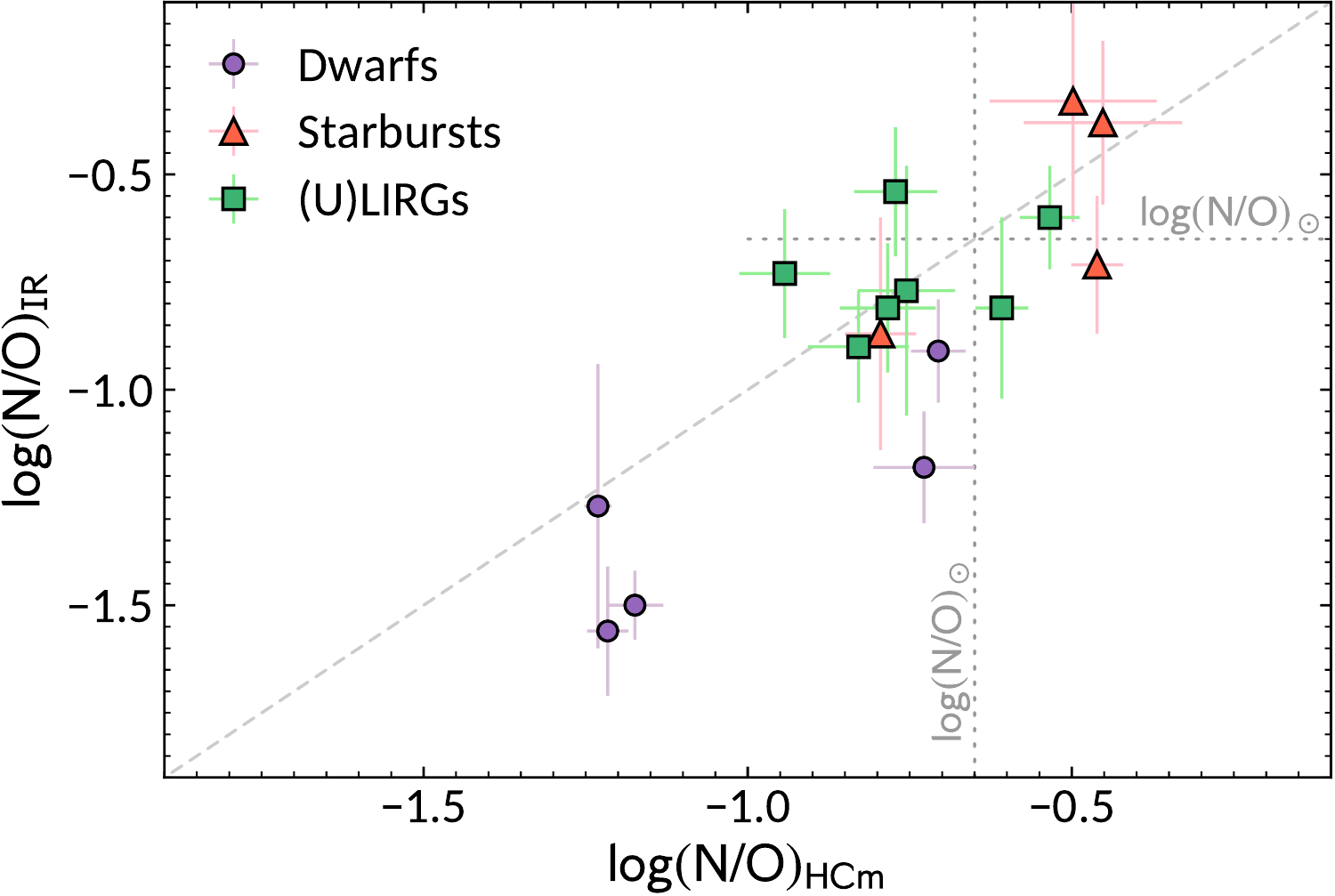}
  \caption{\textbf{Upper left:} comparison between the direct method abundances from \citet{madden2013} and those obtained with \textsc{HCm} \citep{perezmontero2014} using the same optical line flux measurements for the sample of dwarf galaxies (purple circles). \textbf{Upper right:} direct method abundances against the values obtained with \textsc{HCm-IR} using the nebular lines in the IR range. \textbf{Middle left:} comparison between the optical strong-line estimates obtained with the \textit{R} calibration from \citet{pilyugin2016} and \textsc{HCm-IR} for dwarfs galaxies (purple circles), nearby starburst galaxies (red triangles), and local (U)LIRGs (green squares). \textbf{Middle right:} optical abundances obtained with the \textit{S} calibration from \citet{pilyugin2016} against the IR determinations with \textsc{HCm-IR}. \textbf{Lower left:} IR \textsc{HCm-IR} abundances compared to the optical abundances derived with \textsc{HCm}. \textbf{Lower right:} comparison of the N/O abundances obtained with \textsc{HCm-IR} and \textsc{HCm} using the IR and optical nebular lines, respectively, for the galaxies in our sample where both estimates were possible. A linear regression fit between \textsc{HCm-IR} and the optical methods is represented by the black solid line and the associated $95\%$ confidence interval (black dotted lines). In all cases the grey dashed line represents the bisector of the figure. Filled symbols correspond to galaxies where the IR abundance was estimated using a hydrogen recombination line (mostly Hu$\alpha$), while for the galaxies with open symbols the abundances were obtained using only the IR nebular lines.}\label{fig_IRopt}%
\end{figure*}

First of all, we compare the optical \textsc{HCm} and the direct method abundances to show the accuracy of our model-based results using the same observables as in \citet{madden2013}. For the sample of dwarf galaxies, \textsc{HCm} is in excellent agreement with the direct method showing a negligible offset of $-0.02\, \rm{dex}$ and a standard deviation of $0.15\, \rm{dex}$ (upper left panel in Fig.\,\ref{fig_IRopt}). The overall consistency remains when the IR lines are used with \textsc{HCm-IR}, showing an offset of $-0.08\, \rm{dex}$ with a larger scatter of $0.23\, \rm{dex}$ when compared to the direct method (upper right panel; see Table\,\ref{tab_comp2}), mostly due to the apparent saturation of the IR estimates below one tenth of the solar metallicity ($\lesssim 7.6$). Above $7.6$ the average offset decreases to $-0.01\, \rm{dex}$. Fig.\,\ref{fig_IRopt} also shows the linear regression fit in this range between the IR and the direct method measurements (black solid line) with the associated $95\%$ confidence interval (black dotted lines). Most of the values are consistent with the best fit line within $2 \sigma$. A similar scatter between IR and optical abundances has been recently reported by \citet{peng2021} from the N/O abundance measurements derived using the [\ion{N}{iii}]$_{\rm 57 \mu m}$ and [\ion{O}{iii}]$_{\rm 52 \mu m}$ lines compared to the optical estimates based on the [\ion{N}{ii}]$\lambda 6584$ and [\ion{O}{ii}]$\lambda 3727$ lines. The variations found in their study, up to $0.4\, \rm{dex}$ even for unobscured dwarf galaxies such as Haro\,3, are ascribed to the gas stratification within the ISM of these galaxies, since the IR lines should probe a deeper and younger region due to their higher excitation conditions. This is also in line with the results from \citet{polles2019} and \citet{cormier2019} for the same dwarf galaxy sample included in this work, where model-based estimates overpredict the H$\alpha$ to [\ion{O}{iii}]$_{\rm 88 \mu m}$ ratio for nearly half of the sample. These results suggest that the different gas regions probed by IR and optical nebular lines may be a major source of dispersion when \textsc{HCm-IR} estimates are compared to the optical determinations. On the other hand, the higher IR estimates at low metallicities predicted by \textsc{HCm-IR} are likely caused by the large uncertainties that affect the empirical laws in this range, which are used to constrain the grid of photo-ionisation models in \textsc{HCm-IR}. A better observational characterisation of the IR spectrum in extremely low metallicity environments, possible in the near future with the \textit{James Webb Space Telescope} \citep{gardner2006}, will definitely improve the empirical rules inside our code.

For the strong-line methods, the comparison could be extended to the whole sample of galaxies. The colour-filled symbols in Fig.\,\ref{fig_IRopt} correspond to galaxies whose IR-based O/H abundances have been derived using a hydrogen recombination line in the calculation, either Hu$\alpha$ or Br$\alpha$, providing a more robust estimate. The \textit{R} calibration in \citet{pilyugin2016} shows an offset of $-0.11$ to $+0.13\, \rm{dex}$ and a scatter of $0.22\, \rm{dex}$ with respect to \textsc{HCm-IR} (middle left panel in Fig.\,\ref{fig_IRopt}; Table\,\ref{tab_comp2}). Above $12 + \log(O/H) > 7.6$ the maximum offset is seen at the highest metallicities for the sample of starburst galaxies, while the dwarf galaxies show a similar behaviour than in the previous case. A similar result was obtained for the \textit{S} calibration ($-0.12$--$0.13\, \rm{dex}$ offset, $\lesssim 0.24\, \rm{dex}$ standard deviation; middle right panel in Fig.\,\ref{fig_IRopt}), which also reveals the saturation of IR estimates at very low metallicities. As in the previous case, the best fit line follows a slope close to unity while most of the values are consistent within the $2 \sigma$ interval.

A detailed comparison between \textsc{HCm} and \textsc{HCm-IR} allows us to further investigate the possible impact of dust obscuration and the temperature stratification effects on the optical estimates. The differences between the two methods are already shown by the Ne2Ne3 and S3S4 parameters in the right panes of Figs.\,\ref{Ne23}, \ref{S34}, and \ref{N23}. These IR-based parameters present a clear trend of increasing $\log U$ with decreasing optical-based O/H abundances in the star-forming galaxies, although the sulphur and nitrogen parameters seem to be slightly shifted towards lower and higher metallicities, respectively, when compared to the neon parameter. The O3N2 parameter in Fig.\,\ref{O3N2} depends on both $U$ and the relative N/O abundance, although the dependency with the former saturates below $12 + \log(O/H) \lesssim 8.3$--$8.7$. IR vs. optical abundances derived with \textsc{HCm} are compared for the three samples of galaxies in} the lower left panel of Fig.\,\ref{fig_IRopt}. Both methods show consistent results over the $12 + \log(O/H) \sim 7.6$--$8.9$ range within an average offset of $-0.03\, \rm{dex}$ and a standard deviation of $\sim 0.20\, \rm{dex}$ (a factor 1.6) around the bisector, indicated by the grey dashed line in the figure. 
These values are very similar to those found for the comparison with the strong-line calibrations (see Table\,\ref{tab_comp2}), and are also in agreement with those obtained for the dwarf galaxy sample. In our sample, we do not find a significant systematic difference between the optical and IR metallicities as was previously reported by \citet{herrera-camus2018} using the N3O3 tracer from \citet{nagao2011} for a ULIRG sample. The largest average offsets in our starburst and (U)LIRG samples, against the \citet{pilyugin2016} calibrations, are still of the order of the overall scatter, and might be a consequence of the low statistics in those subsamples ($\sim 10$ objects). The individual abundance measurements for each galaxy using the different methods can be found in Table\,\ref{tab_abund}.

In Table\,\ref{tab_comp3} we present the average offset and scatter found between the abundances derived with \textsc{HCm-IR}, using the different line combinations shown in Table\,\ref{tab_comp1}, and the optical-based estimates. We note that the values for the direct method correspond only to the sample of dwarf galaxies. Due to the scarce number of galaxies with [\ion{N}{iii}]$_{\rm 57 \mu m}$ observations from \textit{Herschel}/PACS \citep{jafo2016}, the statistics when this line is required are low ($\lesssim 10$), increasing the average offsets in some cases. Overall, when all the sulphur and neon lines are available, the average offset is typically $\lesssim 0.1\, \rm{dex}$ and the scatter $\sim 0.2\, \rm{dex}$ with respect to any of the optical calibrations adopted here. Including one of the hydrogen recombination lines in the analysis tend to improve these values by slightly decreasing the scatter $\lesssim 0.16\, \rm{dex}$. On the other hand, using only the two sulphur or the two neon lines to estimate the metallicity results in larger offsets ($\sim 0.2$--$0.3\, \rm{dex}$) and/or dispersion ($\sim 0.25\, \rm{dex}$) in all cases. Thus, O/H abundances are optimally derived with \textsc{HCm-IR} when all the sulphur an neon lines are detected, while including the significantly fainter hydrogen recombination lines results in a minor improvement in the scatter with regard to the optical calibrations.

Nevertheless, the nitrogen and oxygen lines are still necessary to determine the N/O ratio. In Fig.\,\ref{fig_IRopt}, lower right panel, we compare the N/O best-fit abundances obtained for the sixteen galaxies where both IR and optical abundances have been determined with \textsc{HCm-IR} and \textsc{HCm}, respectively. For ten of them the N/O abundances obtained are consistent within the errors, three lie still close to the bisector line (in grey) and the remaining three dwarfs show $0.2$--$0.4\, \rm{dex}$ lower N/O abundances in the IR, in line with the results recently found by \citet{peng2021}. Still, this comparison suggests that the N/O abundances obtained using both methods --\,based on the same photo-ionisation simulations but using a completely different set of lines\,-- are consistent for the majority of galaxies.

From the analysis performed in this Section, and considering the intrinsic differences between the optical and IR observables, we conclude that \textsc{HCm-IR} provides reliable O/H and N/O abundance estimates at metallicities above $12 + \log(O/H) \gtrsim 7.6$ with a negligible systematic offset above this limit (e.g. $-0.01\, \rm{dex}$ against the direct method for the dwarf galaxies, in contrast to $-0.08\, \rm{dex}$ for the whole sample), and a typical scatter of about $\sim$\,$0.2\, \rm{dex}$. This is a remarkable result considering that the strong-line calibrations by \citet{pilyugin2016} show, for the sample of dwarf galaxies used in this work, a dispersion of $\sim 0.1$--$0.15\, \rm{dex}$, when compared to the results derived from the same observables using the direct method. Using the IR nebular lines, \textsc{HCm-IR} has demonstrated a slightly higher but comparable accuracy to these calibrations, with a scatter of $0.18\, \rm{dex}$ against the direct method when dwarf galaxies above $12 + \log(O/H) \gtrsim 7.6$ are considered.

\section{Discussion}\label{discuss}

\subsection{Dust obscuration}\label{discuss_dust}
One of the main advantages of IR over optical diagnostics is their low response to dust extinction, which allows to probe the chemical abundances including the dusty embedded star-forming regions hidden to optical wavelengths. Overall, the difference between IR and optical estimates in our sample of galaxies is small. This is expected in dwarf galaxies, since their ISM is poor in heavy elements and dust, indeed the median difference between IR and optical abundances above one tenth of the solar metallicity is negligible ($-0.01\, \rm{dex}$). The scatter in the measurements is likely ascribed to dust obscuration and the different gas regions probed by IR and optical lines due to gas stratification \citep[e.g.][]{dors2013,cormier2019,peng2021}, and the differences between the optical and IR slit apertures, which do not cover exactly the same area in most of these galaxies. In addition to the nebulae seen in the optical, the IR lines also probe the abundances in dust-embedded regions and the cold ionised gas component ($\sim$\,$1\,000\, \rm{K}$) of the ISM, which are not traced by the optical nebular lines (see Fig.\,\ref{fig_emiss}). On the other hand, the temperature dependency may enhance or lessen the contribution of spatially-unresolved regions within galaxies to the optical line fluxes, increasing the scatter when optical and IR estimates are compared. Starburst galaxies show slightly higher IR abundances with a median difference of $0.06\, \rm{dex}$ above the optical values (Fig.\,\ref{fig_IRopt}, lower left panel), although the scatter is larger with a standard deviation of $0.2\, \rm{dex}$, and therefore the difference is not significant. Still, this discrepancy can be important for individual galaxies, e.g. the IR metallicity in NGC\,3198, $12 + \log(O/H)_{\rm IR} = 8.96 \pm 0.05$, is a factor of two larger when compared to the optical metallicity, $12 + \log(O/H)_{\rm opt} = 8.65 \pm 0.03$ (see Fig\,\ref{fig_IRopt}). This is likely caused by obscuration, since NGC\,3198 shows a bright IR nucleus which is not revealed in the UV and optical images \citep{gildepaz2007,brown2014,kennicutt2011}, while a relatively strong absorption is detected in the silicate band at $9.7\, \rm{\micron}$ ($S_{\rm sil} = -1.9$; \citealt{smith2007}). Most of the starburst galaxies in our sample are not heavily absorbed, however, the obscuration in the nebular gas is expected to increase significantly above star formation rates of $\gtrsim 20\, \rm{M_\odot\,yr^{-1}}$, especially at sub-solar metallicities, as has been observed in main-sequence galaxies at $z \sim 3$ \citep{reddy2015,shivaei2020}.

A remarkable result is the similar oxygen abundance found with the optical and the IR methods for the sample of local (U)LIRGs. Both methods show sub-solar metallicities, with only slightly larger values in the IR ($0.07\, \rm{dex}$). (U)LIRGs are typically associated with significant amounts of dust and therefore a difference between the IR and optical metallicities could be expected. This is not the case for the nine (U)LIRGs in our sample, as previously suggested by \citet{pereira-santaella2017} using a calibration of the N3O3 parameter. Our IR abundances are further supported by direct method measurements, since hydrogen recombination lines are detected for half of the (U)LIRG sample. This result suggests that the enriched gas in these local (U)LIRGs is well mixed within the ISM, a scenario that may be supported by the merger-like morphology and interactions typically associated with these galaxies.

\subsection{Comparison with N3O3 based tracers}
Due to the twofold origin of nitrogen, which has a secondary production channel that increases with the overall metallicity, the N3O3 parameter has been proposed as a proxy of the oxygen abundance once a certain calibration of the N/O--O/H relation is assumed \citep{nagao2011}. This diagnostic has been applied to the sample of (U)LIRGs analysed in this work by both \citet{pereira-santaella2017} and \citet{herrera-camus2018} with mismatching results, obtaining IR metallicities in the $0.7 \lesssim Z/Z_\odot \lesssim 1.5$ and $1.5 \lesssim Z/Z_\odot \lesssim 2.5$ ranges, respectively. This discrepancy is mainly caused by a different model calibration, for instance adopting $\log(N/O)_\odot = -0.95$ \citep{charlot2001} results in $\log(N/O) \sim -0.2$ to $0\, \rm{dex}$, a factor of 2 higher when compare to the $\log(N/O)_\odot = -0.65$ calibration \citet{pilyugin2014}. The N/O relative abundances derived with the former calibration are a factor $4$ ($0.6\, \rm{dex}$) higher when compared with our results for the same galaxies using \textsc{HCm-IR}, which are consistent with the N/O ratios derived with \textsc{HCm} using the optical line fluxes for seven (U)LIRGs in our sample (Fig.\,\ref{fig_IRopt}, lower right panel).

This result demonstrates the ability of \textsc{HCm-IR} to provide robust IR-based abundances without relying on a particular calibration of N/O with the overall metallicity. These calibrations can be significantly affected by several factors such as the star forming efficiency, the IMF, and the particular history of gas accretion, star formation, and outflows in each galaxy \citep{molla06}. These factors are typically not well constrained, and thus care must be taken when local calibrations are applied to galaxies with different star formation properties \citep[e.g.][]{amorin10}.



\subsection{Abundances in galaxies at high redshift}\label{highz}
Redshifted far-IR emission lines have been used to probe the ISM of galaxies at high-$z$ using ground-based telescopes in the sub-mm range \citep[e.g.][]{maiolino2005}. One of the most relevant aspects in this regard is the chemical composition of these galaxies, given the critical implications for our understanding of the build-up of heavy elements in the Universe predicted by current models of galaxy evolution \citep[e.g.][]{torrey2019}. So far, the most important metallicity diagnostics at high-$z$ rely on the N3O3 (Eq.\,\ref{eq_n3o3}) and O3N2 parameters (Eq.\,\ref{eq_o3n2}), which are among the brightest nebular lines that can be observed from the ground. We have applied \textsc{HCm-IR} to a sample of eight sub-mm galaxies and Ly$\alpha$ systems at high redshift ($1.8 \lesssim z \lesssim 7.5$) with available measurements in the literature (Table\,\ref{tab_sample}). For three of these galaxies the N3O3 parameter provides a measure of the N/O abundance ($\log(N/O) = -0.8$ to $-1.0\, \rm{dex}$), which is in agreement with the values found in local star-forming galaxies (Fig.\,\ref{n3o3}). For the remaining five galaxies the [\ion{N}{iii}]$57\, \rm{\micron}$ line was not observed, thus the O3N2 parameter is used to estimate the oxygen abundance. All the galaxies show solar-like values ($8.6 \lesssim 12 + \log(O/H) \lesssim 9.0$; orange squares in Fig.\,\ref{O3N2}), in agreement with previous works \citep{ferkinhoff2011,ferkinhoff2015,uzgil2016,rigopoulou2018,tadaki2019,debreuck2019,novak2019}, suggesting that a highly enriched ISM was already present in massive galaxies at high-$z$. Nevertheless, a measurement of the [\ion{N}{iii}]$57\, \rm{\micron}$ line for these galaxies is still necessary to avoid the uncertainties introduced when solar-like N/O ratios are adopted.

Current and future ALMA programmes dedicated to observe the rest-frame IR spectrum of high-$z$ galaxies will soon increase the number of nebular line detections allowing us to provide independent N/O abundances and metalicities for these galaxies using \textsc{HCm-IR}.

\section{Summary}\label{sum}

In this study we present \textsc{Hii-Chi-mistry-IR} (\textsc{HCm-IR}), a new method to derive metallicities from nebular IR lines in star-forming environments. The major advantage of nebular lines in the mid- to far-IR range relies mainly on their weak response to gas temperature and the negligible effect of dust extinction, which seriously affect the optical line intensities. The chemical abundances in the gas-phase are determined by applying bayesian techniques on a grid of photo-ionisation models that covers a wide range in O/H, N/O, and $U$, using known empirical laws between the O/H abundance and the ionisation parameter ($U$) as priors to constrain the grid of models.

The code capabilities are probed using three different galaxy samples: low-metallicity dwarf galaxies, nearby starbursts, and luminous IR galaxies in the local Universe. \textsc{HCm-IR} provides robust IR-based abundances even in the absence of hydrogen recombination lines, with minor variations ($0.03\, \rm{dex}$) within the uncertainty range. When compared with different optical-based determinations using the direct method, photo-ionisation models, and two strong-line calibrations, \textsc{HCm-IR} estimates show a typical scatter of $\sim 0.2\, \rm{dex}$ and do not present systematic differences in the $7.6 < 12 + \log(O/H) < 8.9$ range. This accuracy is reached using only the sulphur and neon lines in the mid-IR, although independent N/O determinations rely on the oxygen and nitrogen lines in the far-IR. These values are slightly higher but comparable to the typical offset and dispersion affecting optical strong-line calibration estimates when compared to the direct method for the sample of dwarf galaxies. Bearing in mind that optical and IR line-emitting regions would likely differ in galaxies due to dust extinction and temperature stratification in the unresolved ISM gas, these results support \textsc{HCm-IR} as a promising new method to exploit the potential of IR spectroscopy to determine chemical abundances. We confirm that obscuration is one of the factors contributing to the scatter between the IR and optical abundances, shown by the case of NGC\,3198. Below $12 + \log(O/H) < 7.6$, \textsc{HCm-IR} abundances tend to saturate due to the large uncertainties that affect the empirical laws in this range, used to link O/H and $U$ and constrain the model grid. Additionally, an important aspect is the independent measurement of the N/O relative abundance that can be obtained when the oxygen and nitrogen lines are detected. The oxygen abundances determined in this case do not rely on the uncertain calibration of this ratio, which may seriously affect ($\sim 0.4\, \rm{dex}$) the metallicity estimates when solar-like N/O ratios are adopted.

We applied \textsc{HCm-IR} to a sample of eight high-$z$ galaxies using nebular far-IR lines redshifted in the sub-millimetre range. The oxygen solar-like abundances obtained for these galaxies, in the $8.6 \lesssim 12 + \log(O/H) \lesssim 9.0$ range, are in agreement with previous studies, although independent N/O measurements based on the N3O3 parameter are necessary to confirm these values.

Nebular IR transitions offer unique diagnostics, independent of temperature and extinction, to measure the abundances of heavy elements in an homogeneous way over a broad range of environments and physical conditions. The same tracers can be used from dusty regions in the centre of nearby galaxies to low-metallicity galaxies at high-$z$. IR metallicity tracers can be currently applied to galactic regions and bright galaxies using the \textit{SOFIA} airborne observatory, and at very high-$z$ when some of these transitions are redshifted into the range accessible to ground-based sub-mm telescopes (ALMA, APEX, CSO). Nevertheless, these tools will be mainly exploited in the future by the new generation of space IR observatories, such as the \textit{James Webb Space Telescope} \citep{gardner2006} in the local Universe. A preliminary research on the exploitation of chemical abundance tracers to study the galaxy evolution discussing possible observational strategies for a future mid- to far-IR spectroscopic observatory \citep[\textit{SPICA};][]{roelfsema2018} was presented in \citet{jafo2017}.

\begin{acknowledgements}
The authors thank the reviewer for his/her comment and suggestions to improve the manuscript. JAFO and LS acknowledge financial support by the Agenzia Spaziale Italiana (ASI) under the research contract 2018-31-HH.0. We acknowledge financial support from the State Agency for Research of the Spanish MCIU through the ``Center of Excellence Severo Ochoa'' award to the Instituto de Astrof\'isica de Andaluc\'ia (SEV-2017-0709). This work has been partly funded by projects ``Estallidos6'' AYA2016-79724-C4 (Spanish Ministerio de Econom\'ia y Competitividad), ``Estallidos7'' PID2019-107408GB-C44 (Spanish Ministerio de Ciencia e Innovaci\'on), and from the Junta de Andaluc\'ia Excellence project EXC/2011 FQM-7058. RA acknowledges support from ANID FONDECYT Regular Grant 1202007. EPM also acknowledges the assistance from his guide dog Rocko without whose daily help this work would have been much more difficult.
\end{acknowledgements}

\bibliographystyle{aa}
\bibliography{HCm-IR}

\end{document}